%% file: main.tex
\documentclass[conference]{IEEEtran}
\IEEEoverridecommandlockouts
\usepackage{cite}
\usepackage{amsmath,amssymb,amsfonts}
\usepackage{xcolor}
\usepackage{listings}
\usepackage{graphicx}
\usepackage{tikz}
\usepackage{cleveref}
\usepackage{url}
\usepackage{pgf}
\usepackage{booktabs}
\usepackage{pgfplots}
\usepackage{algorithm}
\usepackage{algpseudocode}
\pgfplotsset{compat=1.18}

\usepgfplotslibrary{groupplots}
\usepackage{textcomp}
\usepackage{xcolor}
\def\BibTeX{{\rm B\kern-.05em{\sc i\kern-.025em b}\kern-.08em
    T\kern-.1667em\lower.7ex\hbox{E}\kern-.125emX}}
\begin{document}

\title{Compass: Optimizing Compound AI Workflows for Dynamic Adaptation}

\author{
    % \IEEEauthorblockN{Anonymous Authors}
    \IEEEauthorblockN{Milos Gravara}
    \IEEEauthorblockA{
    Distributed Systems Group \\
    TU Wien \\
    m.gravara@dsg.tuwien.ac.at
}
\and
    \IEEEauthorblockN{Juan Luis Herrera}
    \IEEEauthorblockA{
    Distributed Systems Group \\
    TU Wien \\
    j.gonzalez@dsg.tuwien.ac.at
}
\and
\IEEEauthorblockN{Stefan Nastic}
\IEEEauthorblockA{
Distributed Systems Group \\
TU Wien \\
s.nastic@dsg.tuwien.ac.at
}
}

\maketitle

\begin{abstract}
Compound AI is a distributed intelligence approach that represents a unified system orchestrating specialized AI/ML models with engineered software components into AI workflows. Compound AI production deployments must satisfy accuracy, latency, and cost objectives under varying query loads. However, many deployments operate on fixed infrastructure where horizontal scaling is not viable. Existing approaches optimize solely for accuracy and do not consider changes in workload conditions. We observe that compound AI systems can switch between configurations to fit infrastructure capacity, trading accuracy for latency based on current load. This requires discovering multiple Pareto-optimal configurations from a combinatorial search space and determining when to switch between them at runtime. We present Compass, a novel framework that enables dynamic configuration switching through offline optimization and online adaptation. Compass consists of three components: COMPASS-V algorithm for configuration discovery, Planner for switching policy derivation, and Elastico Controller for runtime adaptation. COMPASS-V discovers accuracy-feasible configurations using finite-difference guided search and a combination of hill-climbing and lateral expansion. Planner profiles these configurations on target hardware and derives switching policies using analytical queuing theory based model. Elastico monitors queue depth and switches configurations based on derived thresholds. Across two compound AI workflows, COMPASS-V achieves 100\% recall while reducing configuration evaluations by 57.5\% on average compared to exhaustive search, with efficiency gains reaching 95.3\% at tight accuracy thresholds. Runtime adaptation achieves 90-98\% SLO compliance under dynamic load patterns, improving SLO compliance by 71.6\% over static high-accuracy baselines, while simultaneously improving accuracy by 3-5\% over static fast baselines.
\end{abstract}


\input{sections/introduction}
\input{sections/background_motivation}

\input{sections/compass}
\input{sections/algorithm}
\input{sections/elastico}
\input{sections/evaluation}
\input{sections/related_work}
% \input{sections/limitations}
\input{sections/conclusion}

\section*{Acknowledgment}

This work was partly funded by the European Union under the Horizon Europe programme through the SNS JU (Grant Agreement No. 101192912, NexaSphere). Views expressed are those of the authors and do not necessarily reflect those of the EU or the SNS JU.

\bibliographystyle{IEEEtran} 
\bibliography{references} 

\end{document}

%% file: sections/introduction.tex
\section{Introduction}\label{sec:intro}

The Artificial Intelligence (AI) landscape is evolving from deploying monolithic AI models towards system-centric approaches that utilize and coordinate multiple specialized AI and non-AI components~\cite{compound-ai-blog, Gravara2025ANC}. Known as Compound AI, it represents a distributed intelligence approach combining multiple specialized AI models with tools, and specialized algorithms into workflows orchestrated to solve various AI tasks~\cite{Gravara2025ANC}. This approach offers practical advantages for reliability, scalability and efficiency in AI systems, enabling control over model outputs, component-specific adjustments and optimizations, and adaptation to changing conditions~\cite{chen2023frugalgptuselargelanguage, HybridLLM}. However, Compound AI also introduces complexity. While monolithic models present a single optimization target, Compound AI systems often include many interacting components which affect overall system behavior, each with its own performance characteristics~\cite{OPTSURVEY}. Optimizing and adapting such systems at runtime remains an open research question. 

One of the key challenges in Compound AI is the non-differentiable nature of these workflows~\cite{chen2025llmselector, compound-ai-blog}. Each component in the workflow exposes adjustable parameters, such as hyperparameters and model choices. A Compound AI configuration is one complete setting of these parameters across all components. The variety of parameter choices creates a combinatorial space of possible configurations~\cite{optimas}. Due to heterogeneity of parameter types, gradient-based optimization methods cannot be applied. Instead, finding configurations that optimize a Compound AI workflow for a given task requires searching this discrete configuration space. Black-box optimization techniques can navigate this space, but they require many evaluations to converge, making them impractical for runtime adjustments~\cite{compound-ai-blog}.

Discovering a suitable Compound AI configuration, however, addresses only part of the challenge. Once deployed in production, Compound AI workflows must satisfy Service Level Objectives (SLOs) that span multiple dimensions. These dimensions typically include a) task performance, such as accuracy, b) system performance, such as latency and throughput, and, c) operational costs~\cite{MURAKKAB}. These objectives are competing and introduce trade-offs. Improving accuracy typically demands larger models, which in turn increases latency and cost~\cite{Schahram1}. Traditional cloud-based distributed systems address such trade-offs through elastic horizontal scaling, and theoretically unlimited on-demand resources. This assumption does not hold in many deployment scenarios, including edge infrastructure and dedicated on-premise deployments~\cite{Schahram2, Jellyfish}. In such scenarios, configuration optimized solely for accuracy may require computationally expensive model invocations that limit throughput, causing latency SLO violations when load increases. Recent work in AI and machine learning systems has attempted to address some of these issues, yet these efforts focus predominantly on single, monolithic model inference scenarios~\cite{RAMSIS, INFaaS, CLIPPER}.

Rather than viewing trade-offs merely as constraints to manage, we recognize them as an opportunity for dynamic adaptation. The combinatorial configuration space does not have to be explored in its entirety. SLO constraints, especially those concerning accuracy, define regions of feasibility within the combinatorial configuration space, allowing optimization to target satisfactory configurations, rather than globally optimized ones. Furthermore, deploying larger models typically yields improved accuracy at higher latency and cost, while smaller model variants execute faster with reduced resource requirements, in turn sacrificing some accuracy. This relationship suggests a natural adaptation mechanism for Compound AI workflows~\cite{switching}, as different system states require prioritizing different objectives. During periods of higher load, the system can prioritize throughput to prevent SLO violations, while during lower load, it can prioritize accuracy to maximize quality. Previous work recognized this principle in elastic processes for cloud computing, arguing that systems must jointly reason about resources, quality, and cost rather than treating them as independent concerns~\cite{Schahram1, Schahram2}. We extend this insight into Compound AI inference serving. Instead of only scaling infrastructure horizontally, systems can also adapt vertically by switching between Compound AI configurations. This offers an additional adaptation dimension that operates within fixed infrastructure constraints. 

In this paper, we propose Compass, a novel framework for optimizing Compound AI workflows and effectively managing trade-offs between system performance (e.g., latency, throughput) and task performance (e.g., accuracy). To this end, Compass defines an offline configuration search phase and an online adaptation phase. In the offline phase, Compass searches the combinatorial configuration space under task performance SLO constraints, identifying a set of feasible configurations that satisfy accuracy requirements. These configurations are organized into a Pareto front and encoded with switching policies, derived from a queuing theory based analytical model. In the online phase, a runtime controller monitors load conditions and applies the switching policies to select configurations that maintain SLO compliance as conditions change. This separation allows the computationally expensive search to occur once, while adaptation at runtime is achieved through switching between pre-computed Compound AI configurations.

Our main contributions are as follows: 
\begin{itemize}
    
    \item \textbf{COMPASS-V} - An offline optimization algorithm that 
    efficiently discovers accuracy-feasible Compound AI configurations 
    under SLO constraints. The algorithm employs inverse-distance-weighted finite differences for gradient estimation, progressive budgeting with statistical early stopping, and combines hill-climbing with lateral expansion to navigate the combinatorial configuration space. COMPASS-V achieves 100\% recall in finding ground truth configurations, while reducing evaluations by 57.5\% on average against exhaustive baseline.
    
    \item \textbf{Analytical Queuing-Theory Based Model (AQM)} - A novel analytical model for switching policy construction based on queuing theory. We model the inference serving system as an M/G/1 queue and derive configuration-specific thresholds that determine when to switch between configurations to maintain latency SLO compliance. The model incorporates asymmetric hysteresis to prevent oscillation under fluctuating load. 
    
    \item \textbf{Compass} - A Compound AI optimization framework 
    that integrates COMPASS-V for offline configuration discovery, Planner for deployment-specific policy construction using the analytical model, and an inference serving system with Elastico Controller for runtime adaptation. Elastico applies the AQM-derived switching policies to dynamically select configurations from the Pareto front, maintaining SLO compliance under varying workloads. Elastico improves SLO compliance by 71.6\% compared to high-accuracy baselines, while improving accuracy by 3-5\% over fast baselines.
\end{itemize}

The remainder of this paper proceeds as follows. Background and motivation appear in Section~\ref{sec:motivation}, followed by the Compass framework overview in Section~\ref{sec:compass}. The offline configuration discovery algorithm is detailed in Section~\ref{sec:algorithm}, and the Elastico runtime adaptation mechanism in Section~\ref{sec:elastico}. We evaluate our approach in Section~\ref{sec:evaluation}, followed by related work in Section~\ref{sec:related_work}, and concludes in Section~\ref{sec:conclusion}. 

%% file: sections/background_motivation.tex
\section{Background and Motivation}
\label{sec:motivation}

In this section, we motivate the need for dynamic adaptation in Compound AI serving. We first describe Compound AI systems and the scale of their configuration space in Section~\ref{sub:CAIS}. We then discuss production serving requirements in constrained infrastructure scenarios in Section~\ref{sub:ISS}. This leads to our key insight: \emph{accuracy-latency trade-offs enable runtime adaptation, shifting the optimization goal from finding one optimal configuration to identifying a set of Pareto-optimal configurations}, detailed in Section~\ref{sub:CS}.

\subsection{Compound AI Systems}
\label{sub:CAIS}

Compound AI systems combine multiple specialized AI models with engineered software components to solve various AI tasks~\cite{Gravara2025ANC}. Compound AI workflows typically combine AI model invocations combined with deterministic software, such as data retrieval and processing, input routing, or aggregation algorithms~\cite{RAG, COLLABORATIVEOD, Vate}.  

An example of a Compound AI system is the Retrieval-Augmented Generation (RAG) pipeline~\cite{RAG}. In a standard RAG workflow, the system first passes a query to a retriever function  (e.g., BM25) to fetch relevant context from a knowledge base, typically a vector database. The retriever returns a set of $k$ documents, which are then filtered by the reranker function that assesses their relevance to the query. The filtered documents are fed into a large language model (LLM) to generate the final answer. This design grounds the generation in factual data, reducing hallucinations and improving accuracy. 

Each component in a Compound AI workflow exposes a set of adjustable parameters. These parameters can be categorical (e.g., a set of models for a generator), discrete (e.g., retrieval count $k$), or continuous (e.g., model temperature). A Compound AI configuration is defined as one complete assignment of values to all parameters across all components:

\begin{equation}
    c = (p_1, p_2, \dots, p_n), \quad p_i \in P_i
\end{equation}

The set of valid configurations forms the configuration space $C = P_1 \times P_2 \times \dots \times P_n$. This space grows combinatorially with the number of components and parameters~\cite{chen2025llmselector, optimas}. For our relatively basic RAG example, combining 6 generator models, 5 retrieval-k values, 4 reranker models and 4 reranker-k values yields 480 distinct configurations for a single workflow. Due to heterogeneity of parameter types, Compound AI workflows are non-differentiable, and traditional gradient-based methods cannot be applied. Instead, finding an optimal configuration requires searching this discrete combinatorial space.

\subsection{Compound AI Serving}
\label{sub:ISS}

Deploying a Compound AI system in production requires an inference serving system capable of handling varying query loads while meeting latency requirements. An inference serving system typically consists of a request queue that buffers incoming requests, a workflow executor that processes requests using the active Compound AI configuration, and a controller that manages the system state. Requests arrive at a varying rate determined by user behavior, exhibiting patterns such as diurnal cycles, traffic spikes, and unpredictable bursts~\cite{RAMSIS, InferLine, switching, INFaaS}.

In on-premise deployments and dedicated GPU servers, hardware resources are fixed. Unlike cloud environments where capacity can be provisioned on demand, these settings impose an upper bound on the number of requests the system can process per unit of time~\cite{RAMSIS}. The upper bound depends directly on the active Compound AI configuration. Configurations that use larger, more accurate models require more computation per request, reducing the maximum sustainable throughput. Configurations that use smaller, faster models process requests more quickly, increasing throughput at the cost of reduced task performance.

Under varying load, a single accuracy-optimal configuration is insufficient. When load exceeds the throughput capacity of that configuration, requests accumulate in the queue, and response latency increases, eventually violating latency SLO targets. This creates a trade-off between task performance and system performance. Optimizing for accuracy alone ignores the serving conditions under which the system operates.

\subsection{Configuration Switching}
\label{sub:CS}

The trade-off between accuracy and latency can be exploited as an adaptation mechanism. Rather than committing to a single Compound AI configuration, the system can switch between configurations based on current load. For example, during periods of high load, the system selects faster configurations to prevent SLO violations. During low load, it selects more accurate configurations to maximize task performance.

To explore this adaptation potential, we conducted a preliminary evaluation of a RAG pipeline using the SQuAD 2.0 \cite{squadv2} dataset. We evaluated a subset of  72 distinct configurations, varying components such as the number of retrieved documents ($k$), reranker model and the generator model. Figure~\ref{fig:pareto-insight} visualizes the resulting performance landscape.

% \begin{figure}
%     \centering
%     \includegraphics[width=\linewidth]{figures/plots/motivation/PARETO.pdf}
%     \caption{Pareto front in the RAG workflow running on RTX 4090. Each configuration is presented as a tuple (generator, top-k, reranker, rerank-k)}
%     \label{fig:pareto-insight}
% \end{figure}
\input{figures/plots/motivation/pareto}

Moving between configurations on the Pareto front allows trading small amounts of quality for significant latency gains. As illustrated, switching from the highest quality configuration to an efficient alternative yields a 1.6x reduction in percentile-95 latency with only a 2\% drop in F1 score. 

This observation motivates a strategy of dynamic configuration switching: adapting the system in real-time by moving between different configurations. If a system can identify these Pareto-optimal points and switch between them based on current load, it can maintain service levels within fixed infrastructure limits. However, before switching is possible, it is first necessary to identify which configurations constitute this frontier. This shifts the optimization problem from finding one optimal configuration to finding a set of Pareto-optimal configurations that satisfy minimum accuracy requirement.

%% file: figures/plots/motivation/pareto.tex
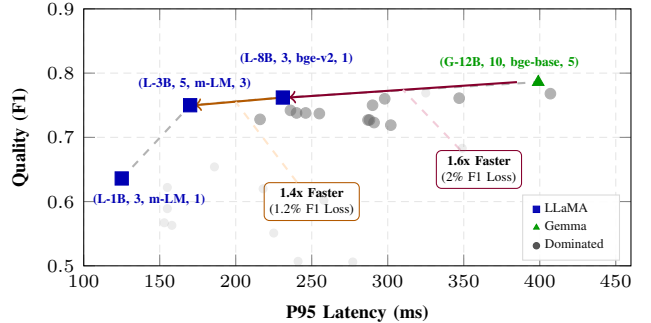
\begin{figure}[h]
\centering
\begin{tikzpicture}
\begin{axis}[
    % Dimensions: Fits single column perfectly
    width=\columnwidth,
    height=5cm,
    % Axis Labels
    xlabel={\textbf{P95 Latency (ms)}},
    ylabel={\textbf{Quality (F1)}},
    label style={font=\scriptsize},
    tick label style={font=\scriptsize},
    % Ranges
    xmin=100, xmax=460,
    ymin=0.5, ymax=0.9,
    % Styling
    grid=major,
    grid style={densely dashed, gray!20},
    % Legend: Bottom Right - Compact
    legend style={
        at={(0.98,0.02)},
        anchor=south east,
        font=\tiny,
        fill=white,
        fill opacity=0.95,
        draw=gray!30,
        inner xsep=4pt,
        inner ysep=1pt,
        row sep=-2pt,
        legend cell align=left,
        /tikz/every even column/.append style={column sep=4pt},
    },
    legend image post style={scale=0.8},
]

% =========================================================
% TIER 3: TOTAL NOISE (Deeply Dominated)
% Low Opacity (0.15) - Just to show the "Search Space"
% =========================================================
\addplot[only marks, mark=*, mark size=1.5pt, color=gray, opacity=0.15, forget plot] 
    coordinates {
    (349, 0.683) (302, 0.411) (241, 0.382) (246, 0.432) 
    (237, 0.467) (258, 0.602) (218, 0.620) (206, 0.449)
    (186, 0.654) (178, 0.431) (155, 0.589) (158, 0.563)
    (155, 0.622) (153, 0.567) (140, 0.359) (225, 0.551)
    (277, 0.506) (241, 0.507) (325, 0.769)
    };

% =========================================================
% TIER 2: NEAR-FRONTIER (The "Almost Good" Configs)
% Medium Opacity (0.5) - Visible, but clearly not winners
% =========================================================
\addplot[only marks, mark=*, mark size=2pt, color=gray!80!black, opacity=0.5, forget plot] 
    coordinates {
    (240, 0.738) (255, 0.737) (290, 0.750)
    (246, 0.738) (216, 0.728)
    (288, 0.727) (291, 0.723) (302, 0.719)
    (236, 0.742) (287, 0.727)
    (347, 0.761) (298, 0.760) (407, 0.768)
    };

% =========================================================
% TIER 1: PARETO FRONTIER (The Heroes)
% Full Opacity, Distinct Colors, Larger Marks
% =========================================================

% The Line
\addplot[thick, dashed, color=gray!60, forget plot] coordinates {
    (125, 0.636) (170, 0.750) (231, 0.762) (399, 0.786)
};

% Legend entries
\addlegendimage{only marks, mark=square*, mark size=1.5pt, color=blue!70!black}
\addlegendentry{LLaMA}
\addlegendimage{only marks, mark=triangle*, mark size=1.5pt, color=green!60!black}
\addlegendentry{Gemma}
\addlegendimage{only marks, mark=*, mark size=1.5pt, color=gray!60!black}
\addlegendentry{Dominated}

% 1. Llama 1B (Fastest)
\addplot[only marks, mark=square*, mark size=2.5pt, color=blue!70!black, forget plot] 
    coordinates {(125, 0.636)};
\node[anchor=north, font=\tiny, color=blue!70!black] 
    at (axis cs:143, 0.626) {\textbf{(L-1B, 3, m-LM, 1)}};

% 2. Llama 3B (Sweet Spot)
\addplot[only marks, mark=square*, mark size=2.5pt, color=blue!70!black, forget plot] 
    coordinates {(170, 0.750)};
\node[anchor=south west, font=\tiny, color=blue!70!black] 
    at (axis cs:130, 0.76) {\textbf{(L-3B, 5, m-LM, 3)}};

% 3. Llama 8B
\addplot[only marks, mark=square*, mark size=2.5pt, color=blue!70!black, forget plot] 
    coordinates {(231, 0.762)};
\node[anchor=north, font=\tiny, color=blue!70!black] 
    at (axis cs:240, 0.846) {\textbf{(L-8B, 3, bge-v2, 1)}};

% 4. Gemma 12B (Max Accuracy)
\addplot[only marks, mark=triangle*, mark size=2.5pt, color=green!60!black, forget plot] 
    coordinates {(399, 0.786)};
\node[anchor=north east, font=\tiny, color=green!60!black] 
    at (axis cs:432, 0.840) {\textbf{(G-12B, 10, bge-base, 5)}};

% =========================================================
% ANNOTATIONS
% =========================================================

\draw[->, thick, color=purple!70!black] (axis cs:385, 0.786) -- (axis cs:235, 0.762);
\node[
    draw=purple!70!black,
    fill=white,
    rounded corners=2pt,
    inner sep=2pt,
    font=\tiny,
    align=center
] (calloutA) at (axis cs:360, 0.65) {%
    \textbf{1.6x Faster}\\
    (2\% F1 Loss)
};

\draw[-, thick, dashed, color=purple!20]
    (calloutA) -- (axis cs:310, 0.774);

\draw[->, thick, color=orange!70!black] (axis cs:231, 0.762) -- (axis cs:173, 0.750);
\node[
    draw=orange!70!black,
    fill=white,
    rounded corners=2pt,
    inner sep=2pt,
    font=\tiny,
    align=center
] (calloutB) at (axis cs:250, 0.6) {%
    \textbf{1.4x Faster}\\
    (1.2\% F1 Loss)
};

\draw[-, thick, dashed, color=orange!20]
    (calloutB) -- (axis cs:200, 0.755);

\end{axis}
\end{tikzpicture}
\caption{Pareto front in the RAG workflow running on RTX 4090. Each configuration is presented as a tuple (generator, top-k, reranker, rerank-k)}
\label{fig:pareto-insight}
\end{figure}

%% file: sections/compass.tex
\section{Compass Overview}
\label{sec:compass}

\begin{figure*}[t]
  \centering
  \includegraphics[width=\linewidth]{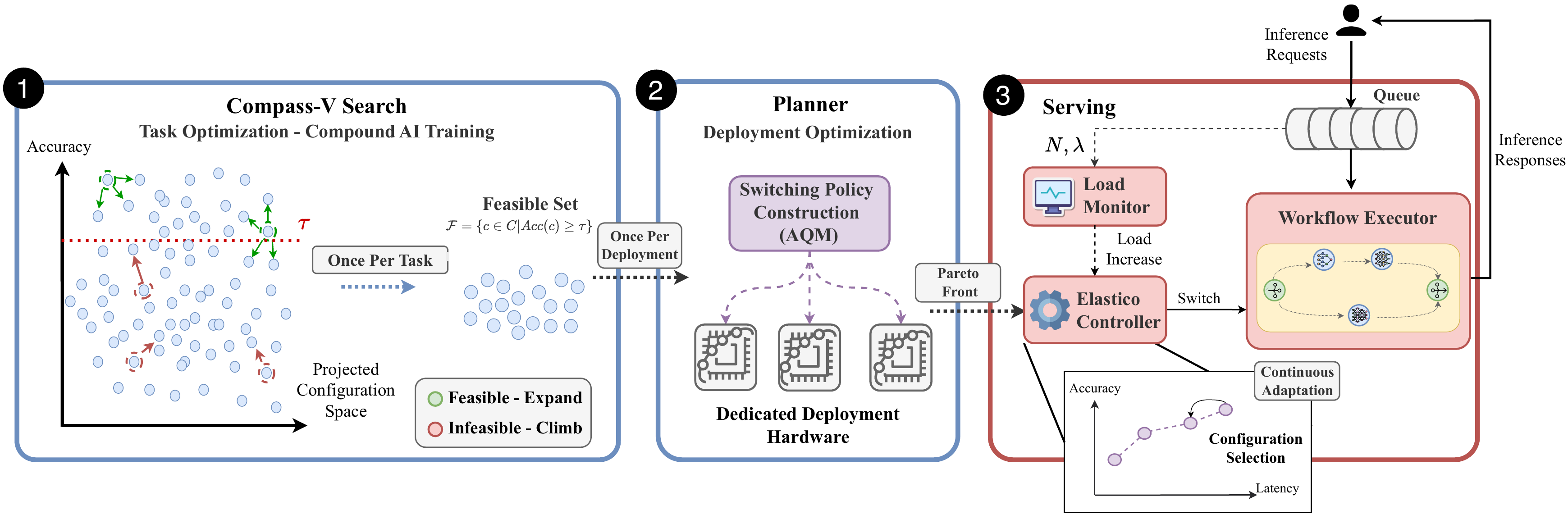}
  \caption{Compass Framework Overview - Offline phase includes task optimization and deployment planning. Online phase represents inference serving system architecture which includes proposed Elastico Controller for dynamic adaptation.}
  \label{fig:compass}
\end{figure*}

We provide a high-level overview of the main system components in Compass (Fig.~\ref{fig:compass}), a novel framework that enables runtime adaptation of Compound AI workflows through configuration switching. The framework is divided into an offline phase and an online phase. The offline phase optimizes the Compound AI workflow for a given task and constructs switching policies based on target deployment hardware. The online phase serves inference requests and adapts the active Compound AI configuration in response to load changes.

Compass requires the following inputs. A Compound AI workflow $W$, where the system designer specifies which component parameters are used for adaptation along with their valid values. This determines the configuration space $\mathcal{C}$. Next, Compass expects a sample dataset $\mathcal{D}$ representing the target task, used for accuracy evaluation during optimization and for latency profiling during planning. This is coupled with an accuracy threshold $\tau$ defining minimum acceptable task performance, a latency SLO $\delta$ specifying maximum acceptable end-to-end response time and a description of the target deployment hardware H.

\subsection{Offline Phase - Optimization and Planning}

The offline phase consists of two stages: task optimization and deployment planning.

Task optimization takes as input the workflow $W$, dataset $\mathcal{D}$, and accuracy threshold $\tau$. Since runtime adaptation requires switching between multiple configurations, Compass formulates task optimization as a search over the configuration space $\mathcal{C}$ to identify all configurations whose evaluated accuracy on $\mathcal{D}$ exceeds $\tau$. The result of this process is the feasible set $\mathcal{F}$, where $|\mathcal{F}| \ll |\mathcal{C}|$, thereby reducing the configuration space that subsequent deployment planning must consider. Task optimization is executed once per task and is independent of the deployment hardware. As a result, the feasible set can be reused across deployments on different infrastructure. To achieve this, we develop \textsc{Compass-V}, a search algorithm that explores the configuration space using progressive budgeting with statistical early stopping and gradient-guided navigation toward feasible regions. A detailed description of the algorithm is provided in Section~\ref{sec:algorithm}.

Deployment planning prepares the feasible set for serving on target deployment hardware. Planner takes as input the feasible set $\mathcal{F}$, target hardware $H$ and latency SLO $\delta$. Building on the prior work~\cite{RAMSIS, Jellyfish, InferLine}, Planner profiles each configuration in $\mathcal{F}$ on the target hardware, recording latency statistics across multiple runs using representative inputs from $\mathcal{D}$. For workflows containing LLM components, latency varies with input and output length, requiring percentile-based profiles to capture this variability. For traditional ML components where inference time is more predictable, mean latency suffices~\cite{InferLine}. From these profiles, Planner finds Pareto-optimal configurations over accuracy and latency, discarding configurations that are dominated on both dimensions. For each Pareto-optimal configuration, Planner uses the analytical queuing theory based model (AQM) to compute a switching threshold that determines the maximum load under which that configuration can meet the latency SLO. The output is a Pareto front: an ordered set of configurations with their accuracy, latency profiles, and switching policies. Since deployment planning depends only on target hardware, if the system is deployed on new infrastructure, only this stage must be re-executed. Section~\ref{sec:elastico} describes AQM and threshold derivation in detail.

\subsection{Online Phase - Runtime Adaptation}

The online phase performs runtime adaptation, adjusting the active configuration in response to changing load. This phase utilizes the Pareto front received from the Planner to switch configurations during serving. The runtime architecture follows established inference serving system architecture~\cite{CLIPPER, RAMSIS, Jellyfish}, and consists of four components: a central queue that buffers incoming inference requests, a load monitor that tracks current queue depth and arrival rate, the Elastico Controller that makes configuration selection decisions, and a workflow executor that processes requests using the active configuration.

Elastico is a runtime controller that uses the load monitor's measurements together with the precomputed switching thresholds from the Pareto front to determine adaptation decisions. When queue depth exceeds the threshold of the current configuration, Elastico selects a faster configuration to sustain the increased load. When queue depth drops below the threshold of a more accurate configuration, it switches to improve accuracy. To prevent oscillation under fluctuating load, Elastico applies asymmetric hysteresis, reacting quickly to load increases but requiring sustained low load before switching to more accurate configurations. During a switch, the executor continues serving requests with the current configuration until the new configuration is ready, ensuring no requests are dropped.

%% file: sections/algorithm.tex
\section{COMPASS-V: Feasible Configuration Search}
\label{sec:algorithm}

This section presents COMPASS-V, the task optimization algorithm in Compass. We first formulate the optimization problem in \cref{subsec:formulation}, then describe the algorithm in \cref{subsec:algorithm} and assess its properties in \cref{subsec:prop}.

\subsection{Problem Formulation}\label{subsec:formulation}

We formulate Compound AI task optimization as a modified hyperparameter optimization problem. Standard hyperparameter optimization focuses on finding a single accuracy-optimal configuration $c^* = argmax_{c\in\mathcal{C}} Acc(c)$. However, runtime adaptation requires multiple configurations to switch between as load conditions change while preserving minimal accuracy threshold. Thus, we reformulate the optimization problem as finding the feasible set:  

\begin{equation}
    \mathcal{F} = \{(c, Acc(c)): c \in \mathcal{C}, Acc(c) \geq \tau\}
\end{equation}

where $\tau$ is an operator-specified accuracy threshold defining minimum acceptable task performance. This formulation shifts the objective from convergence to a single optimum to coverage of feasible region within the configuration space $\mathcal{C}$.

\subsection{COMPASS-V Algorithm}\label{subsec:algorithm}

COMPASS-V navigates the configuration space using estimated gradients
and adaptive exploration. Since Compound AI workflows are non-differentiable,
we estimate gradients via inverse-distance-weighted finite differences
from the $k$ nearest evaluated configurations (Line~\ref{line:idw}--\ref{line:hillclimb}). In other words, COMPASS-V interpolates accuracy differences from $k$ nearest neighbors, weighted by inverse distance:

\begin{equation}
    v_i(c) = \frac{\sum_{n \in N_k(c)} w_n \cdot \frac{\Delta Acc_n}{\Delta x_i}}
            {\sum_{n \in N_k(c)} w_n},
\quad w_n = d(c,n)^{-p}
\end{equation}

where all parameters are normalized to $[0,1]$ to enable distance
computation across heterogeneous types. 

The navigation strategy depends on feasibility status:
\begin{itemize}
  \item \textbf{Hill-climbing} ($Acc(c) < \tau$): Follow the
        gradient toward higher accuracy until reaching the feasible region (Lines~\ref{line:idw}--\ref{line:hillclimb}).
  \item \textbf{Lateral expansion} ($Acc(c) \geq \tau$): Explore
        neighboring configurations along low-gradient axes to trace
        the feasible boundary (Line~\ref{line:lateral}).
\end{itemize}

\textbf{Initialization.} Hill-climbing can become trapped in local optima. To avoid this, COMPASS-V initializes with Latin Hypercube Sampling~\cite{LHS} and evaluates sampled configurations. This enables diverse configuration space coverage and provides initial gradient estimates (Line~\ref{line:lhs}). 

\textbf{Progressive Evaluation.} Since accuracy evaluation requires potentially long-running workflow executions, COMPASS-V uses progressive budgeting: starting with few samples of the budget and increasing only when feasibility remains uncertain. This enables early stopping for configurations that are clearly feasible or unfeasible. However, evaluating on fewer samples can introduce uncertainty. Thus, COMPASS-V utilizes Wilson confidence intervals to quantify this uncertainty. A configuration is classified as feasible only when the lower bound exceeds $\tau$, and infeasible only when the upper bound falls below it. Borderline cases receive additional samples until confident classification (Lines~\ref{line:progstart}--\ref{line:progend}).

\begin{algorithm}[h]
  \caption{COMPASS-V: Feasible Configuration Search}
  \label{alg:compassv}
  \small
  \begin{algorithmic}[1]
  \Require Configuration space $\mathcal{C}$, threshold $\tau$, budget schedule $\{b_1, \dots, b_K\}$
  \Ensure Feasible set $\mathcal{F}$
  
  \State $\mathcal{F} \gets \emptyset$; \; $E \gets \emptyset$ 
  \State $Q \gets \Call{LHSSample}{\mathcal{C},\, n_{\text{init}}}$ 
  \label{line:lhs}
  \Comment{Initialize with diverse samples}
  
  \While{$Q \neq \emptyset$}
      \State $c \gets Q.\text{pop}()$
      
      \Comment{Progressive evaluation with early stopping}
      \For{$b \in \{b_1, \dots, b_K\}$} \label{line:progstart}
          \State $\hat{a},\, [\text{CI}_{\text{lo}},\, \text{CI}_{\text{hi}}] \gets \Call{Evaluate}{c,\, b}$
          \If{$\text{CI}_{\text{lo}} > \tau$  \textbf{or}  $\text{CI}_{\text{hi}} < \tau$}
              \State \textbf{break}
              \Comment{Confident classification}
          \EndIf
      \EndFor \label{line:progend}
      
      \State $E \gets E \cup \{(c,\, \hat{a})\}$
      
      \Comment{Navigate based on feasibility}
      \If{$\hat{a} \geq \tau$}
          \State $\mathcal{F} \gets \mathcal{F} \cup \{(c,\, \hat{a})\}$
          \State $Q \gets Q \cup \Call{LateralExpand}{c,\, E}$ 
          \label{line:lateral}
      \Else
          \State $\mathbf{v} \gets \Call{IDWGradient}{c,\, E}$
          \label{line:idw} \Comment{Eq.~3}
          \State $Q \gets Q \cup \Call{HillClimb}{c,\, \mathbf{v}}$ 
          \label{line:hillclimb}
      \EndIf
  \EndWhile
  
  \State \Return $\mathcal{F}$
  \end{algorithmic}
\end{algorithm}

\subsection{Theoretical Properties}\label{subsec:prop}

COMPASS-V operates over the finite configuration space $\mathcal{C}$ and evaluates each configuration at most once. We summarize the formal properties that follow from this structure.

\textbf{Termination and worst-case complexity.} At each iteration, exactly one configuration is removed from the queue $Q$ and added to the evaluated set $E$. Since no configuration is evaluated twice and $|\mathcal{C}|$ is finite, the algorithm terminates after at most $|\mathcal{C}|$ iterations. Each configuration consumes at most $B_{max} = b_K$ samples through progressive budgeting, yielding a worst-case complexity of $\mathcal{O}(|\mathcal{C}|\cdot B_{max})$. This worst case arises when most configurations concentrate near $\tau$, preventing both early stopping and efficient navigation.

\textbf{Convergence.} For any configuration $c$ with true accuracy $p$, Wilson confidence interval at budget $b_i$ with confidence level $1 - \alpha$ satisfies $P(p \in [CI_{lower}, CI_{upper}]) \ge 1 - \alpha$. As evaluation budget increases, the interval width shrinks, and the classification converges to the correct decision. 

\textbf{Completeness.} We define two configurations as \emph{adjacent} if they differ in exactly one parameter value, inducing a graph over $\mathcal{C}$. If the feasible region $\mathcal{F}$ is connected in this graph and all neighbors are explored at each expansion step, discovering any feasible configuration $c_f \in \mathcal{F}$ triggers breadth-first expansion that discovers all of $\mathcal{F}$. Completeness then reduces to the seeding problem: for $n_{\text{init}}$ Latin Hypercube Sampling (LHS) samples over a space with feasible fraction $f = \frac{|\mathcal{F}|}{|\mathcal{C}|}$, the seeding probability is $P_{\text{seed}} \ge 1 - (1 - f)^{n_{\text{init}}}$. When the feasible region is disconnected, lateral expansion from one feasible configuration cannot reach all feasible configurations. The LHS sampling must cover each disconnected feasible region separately, requiring a larger number of initial samples. 

%% file: sections/elastico.tex
\section{AQM: Analytical Queuing-theory based Model for Runtime Adaptation}
\label{sec:elastico}

This section presents the queuing theory based analytical model for switching policy construction. The planning phase takes the feasible set F from task optimization, profiles latency on target hardware, constructs a Pareto front, and derives switching policies. Elastico then uses these policies at runtime to adapt the active configuration in response to load changes.

\subsection{System Model}

  We model the inference server as an M/G/1 queue with configurations
  $\mathcal{C} = \{c_0, \ldots, c_n\}$ from the Pareto front ordered by
  increasing service time:
  \begin{equation}
      \bar{s}_0 < \bar{s}_1 < \cdots < \bar{s}_n
  \end{equation}
  where $\bar{s}_k = \mathbb{E}[S_k]$ is the mean service time of configuration
  $c_k$. Since latency and accuracy trade off on the Pareto front, this
  implies $a_0 < a_1 < \cdots < a_n$ for accuracies.

  Each configuration has empirical tail latency $s_{95,k}$ measured during
  profiling. Requests are assumed to arrive as a Poisson process with rate $\lambda$ and are served FIFO without preemption.

  \subsection{Response Time Decomposition}

  A request arriving that finds $N$ requests in the system experiences response time:
  \begin{equation}
      T = W + S_k = \sum_{j=1}^{N} S_j + S_k
  \end{equation}
  where $W$ is waiting time (sum of remaining service times of queued requests)
  and $S_k$ is the request's own service time under configuration $c_k$.

  \subsection{SLO Constraint}

  Given a P95 latency SLO target $L$, we require:
  \begin{equation}
    T_{95} \leq L \iff \mathbb{P}(T \leq L) \geq 0.95
  \end{equation}

  We decompose this constraint by allocating the latency budget between
  waiting and service. Reserving the tail service latency for the service
  phase, the waiting time must satisfy:
  \begin{equation}
      W_{95} \leq L - s_{95,k} \triangleq \Delta_k
  \end{equation}
  where $\Delta_k$ is the \emph{queuing slack}: the time budget available
  for waiting. Configurations with $\Delta_k \leq 0$ cannot satisfy the SLO
  and are excluded.

  \subsection{Upscale Threshold}

  For $N$ requests ahead in the queue, each with service time drawn from
  distribution $S_k$, the expected waiting time is:
  \begin{equation}
      \mathbb{E}[W] = N \cdot \bar{s}_k
  \end{equation}

  Using the mean as a proxy for the P95 (exact for deterministic service,
  approximate otherwise), we impose:
  \begin{equation}
      N \cdot \bar{s}_k \leq \Delta_k
  \end{equation}

  Solving for the maximum safe queue depth yields the \textbf{upscale threshold}:
  \begin{equation}
      N_k^{\uparrow} = \left\lfloor \frac{\Delta_k}{\bar{s}_k} \right\rfloor
      = \left\lfloor \frac{L - s_{95,k}}{\bar{s}_k} \right\rfloor
  \end{equation}

  When $N > N_k^{\uparrow}$, the current configuration risks SLO violations
  and the system should switch to a faster configuration $c_{k-1}$.

From the configuration ordering, faster configurations tolerate larger queues:
  \begin{equation}
      N_0^{\uparrow} > N_1^{\uparrow} > \cdots > N_n^{\uparrow}
  \end{equation}

  This creates a ladder: under increasing load, the system progressively
  switches to faster configurations; under decreasing load, it climbs back to
  more accurate ones.

  \subsection{Downscale Threshold}

  To switch from current configuration $c_k$ to a slower but more accurate
  configuration $c_{k+1}$, we must ensure the slower configuration can handle
  the current queue without violating SLOs.

  The downscale condition requires:
  \begin{equation}
      N \cdot \bar{s}_{k+1} \leq \Delta_{k+1} - h_s
  \end{equation}
  where $h_s$ is a \emph{slack buffer} (in milliseconds) that provides margin
  for the transition. This yields the \textbf{downscale threshold}:
  \begin{equation}
      N_k^{\downarrow} = \left\lfloor \frac{\Delta_{k+1} - h_s}{\bar{s}_{k+1}} \right\rfloor
  \end{equation}

  The system downscales to $c_{k+1}$ when $N < N_k^{\downarrow}$, recovering
  accuracy when load permits. 

  \subsection{Asymmetric Temporal Hysteresis}

  Queue-based thresholds alone can cause oscillation under fluctuating load.
  To prevent this, we introduce asymmetric cooldown periods that reflect the different
  costs of each switching direction:

  \textbf{Upscale cooldown} $t^{\uparrow}$: Set to zero or near-zero.
  When the arrival rate spikes, SLO violations are immediate under current configuration. Thus, the system must react quickly to load spikes.

  \textbf{Downscale cooldown} $t^{\downarrow}$: Set to several seconds
  (e.g., 5s). Switching to a slower, but more accurate configuration, too eagerly risks oscillation and unnecessary accuracy loss if load rebounds. The system waits for sustained low load before recovering for accuracy.

  % The switching conditions become:
  % \begin{align}
  %     \text{Upscale:} \quad & N > N_k^{\uparrow} \;\land\; (t - t_{\text{last}}^{\uparrow}) > \tau^{\uparrow} \\
  %     \text{Downscale:} \quad & N < N_k^{\downarrow} \;\land\; (t - t_{\text{last}}^{\downarrow}) > \tau^{\downarrow}
  % \end{align}

Under the M/G/1 model with FIFO scheduling, the derived thresholds
guarantee that queued requests remain within the latency slack, enabling quick reaction of the serving system to the load increase. Asymmetric hysteresis prevents repeated switching and guarantees convergence to the highest-accuracy configuration under lower load. 

%% file: sections/evaluation.tex
\section{Evaluation}
\label{sec:evaluation}

This section evaluates the effectiveness of Compass in supporting both efficient configuration discovery and adaptive runtime execution in Compound AI workflows. Compass is published as an open-source framework within the Polaris project. It is implemented in Python and available on GitHub.\footnote{\url{https://github.com/polaris-slo-cloud/compass}}

First, we evaluate COMPASS-V, measuring its ability to discover all feasible configurations across varying accuracy thresholds with fewer evaluations than exhaustive search. Then, we evaluate Elastico as a runtime adaptation mechanism, examining how dynamic configuration switching under changing load conditions balances accuracy and latency while maintaining SLO compliance.

\subsection{Experimental Setup}

We evaluate Compass on a dedicated inference server with NVIDIA RTX 4090 (24GB VRAM), 32-core CPU, and 128GB RAM running Ubuntu 22.04. Each workflow component executes in a Docker container with NVIDIA runtime, ensuring reproducible resource allocation. For runtime adaptation experiments, all evaluated configurations are pre-loaded in GPU memory (total: 18GB), eliminating cold-start overhead during configuration switches.

\subsection{Compass-V Evaluation}

We evaluate COMPASS-V on two compound AI workflows with different topologies and parameter types. The first is the RAG pipeline introduced in \Cref{sec:motivation}, evaluated on SQuAD 2.0~\cite{squadv2} using F1. The configuration space of 234 configurations is constructed from 6 generator models (LLaMA3:\{1B,3B,8B\} and Gemma3:\{1B,4B,12B\}), 5 retriever-k values (3, 5, 10, 20, 50), 4 reranker-k values (1, 3, 5, 10), and 3 reranker models (BGE-v2, BGE-base, MS-MARCO). 

The second is a multi-model object detection cascade evaluated on COCO~\cite{COCO} using mAP@0.5. In this workflow, a lightweight detector processes every input image. When the detector's confidence falls below a threshold, the prediction is forwarded to a heavier verifier model for re-evaluation. The configuration space of 385 configurations spans 3 detector models (YOLOv8n/s/m), 4 verifier models (YOLOv8m/l/x or none), 7 confidence thresholds (0.1 to 0.5), and 5 NMS thresholds (0.3 to 0.7).

For both workflows, we establish ground truth by exhaustively evaluating all configurations via grid search. COMPASS-V uses a maximum evaluation budget of 100 samples for RAG and 200 samples for object detection. We test 8 SLO thresholds for RAG (0.30 to 0.9) and 8 for object detection (0.55 to 0.8), spanning a range of feasible fractions from 99\% to 2\%. 

We analyze two aspects of the approach: (1) how quickly and accurately COMPASS-V converges to the complete feasible set compared to exhaustive search and (2) overall sample efficiency across the SLO spectrum.

\subsubsection{Convergence Analysis}

\begin{figure*}[!t]
    \centering
    \resizebox{\linewidth}{!}{%
        \input{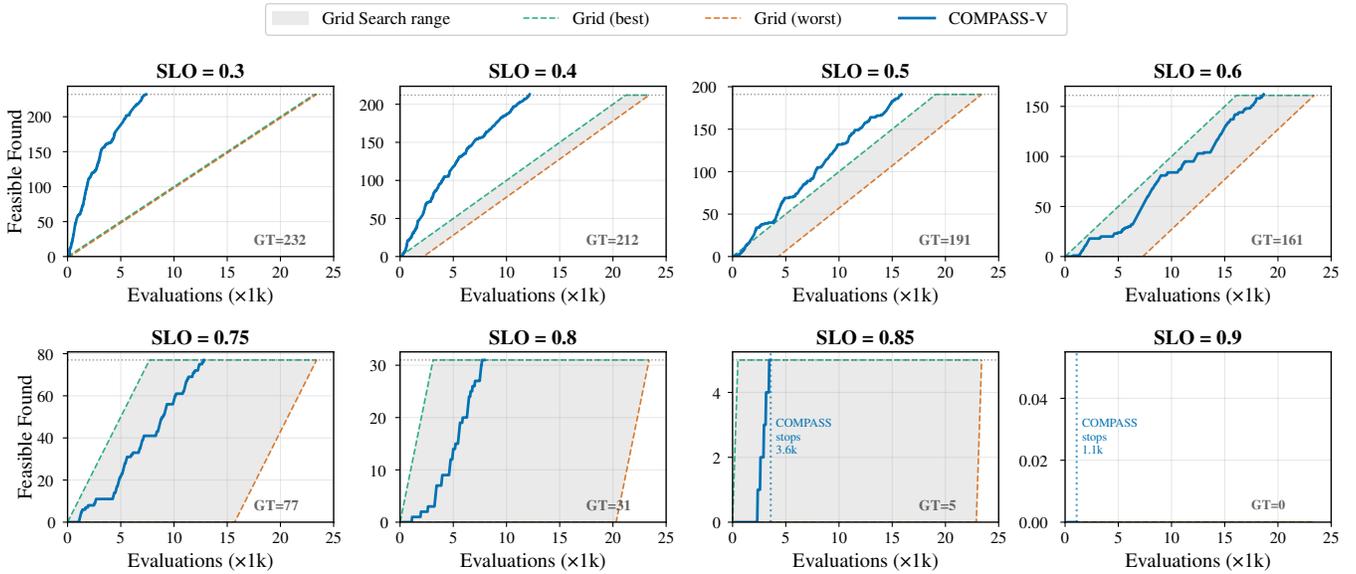}
    }
    \caption{COMPASS-V convergence across various accuracy SLOs.}
    \label{fig:convergence}
\end{figure*}

Figure~\ref{fig:convergence} presents COMPASS-V's anytime convergence behavior across all eight SLO thresholds on the RAG pipeline. Each subplot shows feasible configurations discovered versus sample evaluations consumed. The shaded region represents Grid Search's theoretical range: from best-case (evaluating feasible configurations first) to worst-case (evaluating them last). At SLO 0.5, where 82\% of configurations are feasible, COMPASS-V discovers all 191 feasible configurations through rapid lateral expansion, saturating well before exhaustive search would complete. At SLO 0.75 (33\% feasible), discovery is more gradual as the algorithm navigates a larger infeasible region through hill-climbing before expanding the feasible boundary. At SLO 0.85 (2\% feasible), COMPASS-V locates all 5 feasible configurations and terminates early after exploring only a small fraction of the space. Across all thresholds and both workflows, COMPASS-V achieves 100\% recall compared to exhaustive ground truth.

\begin{figure}[h]
    \centering
    \resizebox{0.8\columnwidth}{!}{%
        \input{figures/plots/experiments/compass_v/fig4_efficiency_scatter.pgf}
    }
    \caption{COMPASS-V efficiency across different accuracy SLOs.}
    \label{fig:efficiency}
\end{figure}
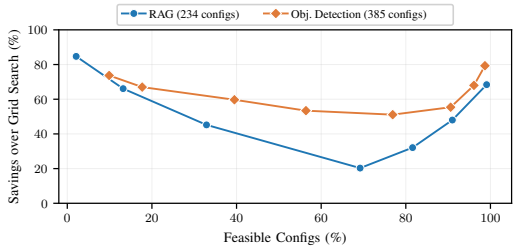

\subsubsection{Sample Efficiency Analysis}

Figure~\ref{fig:efficiency} presents COMPASS-V's evaluation savings against feasible fraction for both workflows. The x-axis shows the percentage of configurations that are feasible at a given SLO threshold, and the y-axis shows the percentage reduction in sample evaluations compared to exhaustive grid search. COMPASS-V achieves 100\% recall across all 16 tested SLO thresholds spanning both workflows. Both workflows exhibit a convex efficiency pattern. At low feasible fractions (left side), most configurations are infeasible, and Wilson CI early stopping classifies them quickly without full budget evaluation. At high feasible fractions (right side), most configurations are clearly feasible, and early stopping resolves them equally fast. Savings are lowest at moderate feasible fractions (60--75\%), where the boundary between feasible and infeasible regions is proportionally largest, requiring more extensive exploration before confident classification. On RAG pipeline, COMPASS-V achieves savings between 20.3\% and 84.7\%, while on the object detection pipeline it achieves savings between 51.1\% and 79.3\%. Both workflows follow the same efficiency trend despite differing in pipeline topology, parameter types, and configuration space size, suggesting that COMPASS-V's efficiency is driven primarily by the feasible fraction rather than workflow-specific properties.

\subsection{Runtime Adaptation}

This subsection evaluates Elastico's ability to dynamically adapt through configuration selection under varying load conditions. We demonstrate that Elastico achieves near-optimal trade-offs between SLO compliance and accuracy by switching between configurations in response to load changes, outperforming static baseline approaches that use fixed configurations.

To stress-test the adaptation mechanism, and similarly to previous work \cite{Jellyfish, InferLine}, we evaluate Elastico under two representative load patterns: (1) \textbf{Spike Pattern} with a sustained 4$\times$ load increase during middle third of the experiment, and (2) \textbf{Bursty Pattern} with random short 2-5$\times$ bursts of high load throughout experiment, lasting 5-15s. 

We scale these patterns to our hardware capacity (base=1.5 QPS) while preserving the relative stress characteristics. The key evaluation metric is not absolute QPS but the system's ability to maintain SLO compliance across load transitions. Each pattern runs for 180 seconds with base QPS of 1.5 requests/second. We test three SLO targets: 500ms ($\sim$slowest configuration), 1000ms ($\sim$1.5x slowest configuration), and 1500ms ($\sim$2x slowest configuration).

We compare Elastico against three static baselines \footnote{Previous work \cite{RAMSIS, Jellyfish, InferLine} targets single-model selection per query. Adapting them to Compound AI would require per-component model selection. This would assume components that are independent, which is not the case for Compound AI workflows.} (Table~\ref{tab:baselines}) that are on the resulting Pareto front constructed through COMPASS-V search for $\tau = 0.75$ and deployment planning phase. All Pareto-optimal configurations are pre-loaded in GPU memory, requiring 18GB total. Configuration switches execute in $<$ 10ms by changing inference pipeline routing, avoiding model loading overhead.

\begin{table}[t]
    \centering
    \caption{Baseline configurations of the generated Pareto front.}
    \scriptsize
    \label{tab:baselines}
    \begin{tabular}{l c c c}
        \hline
        Name & Config (gen, reranker, top-k, rerank-k) & Accuracy (F1) & $P_{95}$ Lat \\
        \hline
        Fast     & (Llama3.2:3B, ms-marco, 20, 1) & 0.761 & $\sim$200\,ms \\
        Medium   & (Llama3.1:8B, ms-marco, 10, 3) & 0.825 & $\sim$450\,ms \\
        Accurate & (Gemma3:12B, bge-v2, 20, 3) & 0.853 & $\sim$700\,ms \\
        \hline
    \end{tabular}
\end{table}

\subsubsection{Accuracy-Latency trade-off}

\begin{figure}[h]
    \centering
    \resizebox{\columnwidth}{!}{%
        \input{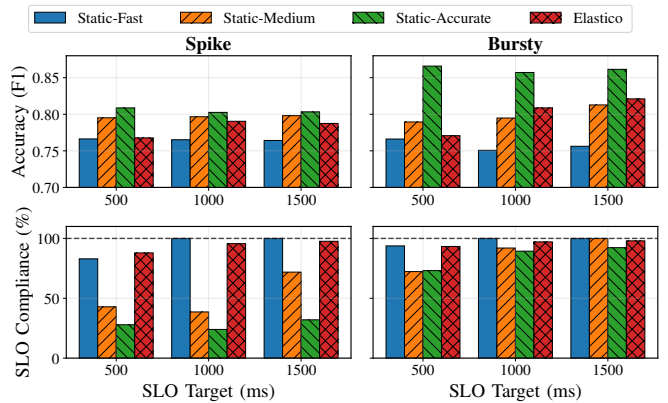}
    }
    \caption{Elastico performance with varying SLO constraints.}
    \label{fig:bars}
\end{figure}

Figure~\ref{fig:bars} reveals the accuracy-latency trade-off that static approaches face. Static-Fast achieves high SLO compliance (82-100\%) but sacrifices accuracy, while Static-Accurate maximizes accuracy but suffers severe SLO compliance degradation under load, dropping to just 24\% compliance during spike conditions at 1000ms SLO. Elastico utilizes this trade-off by switching dynamically between configurations. Under the spike pattern with 1000ms SLO, Elastico improves compliance by 71.6 \% over Static-Accurate while simultaneously improving accuracy by 2.9 percentage points over Static-Fast. This demonstrates that adaptive configuration selection achieves results that single static configurations cannot match.

% The scatter plots (Fig~\ref{fig:scatter}) position each approach in the accuracy-compliance space across all SLO levels. In both scenarios, Elastico occupies the upper-right region representing optimal trade-offs. Static-Fast clusters in the high-compliance, low-accuracy region, while Static-Accurate appears in the high-accuracy, low-compliance region.

% \begin{figure}[h]
%     \centering
%     \resizebox{\columnwidth}{!}{%
%         \input{figures/plots/experiments/elastico/combined_scatter_1.pgf}
%     }
%     \caption{Elastico accuracy latency trade-off.}
%     \label{fig:scatter}
% \end{figure}

% Notably, Elastico's position shifts based on SLO tightness: under tight SLOs (500ms), it prioritizes compliance; under relaxed SLOs (1500ms), it shifts toward higher accuracy. This adaptive behavior demonstrates that Elastico effectively exploits available latency slack to maximize accuracy without compromising SLO guarantees.

\subsubsection{Latency Distribution Analysis}

\begin{figure}[h]
    \centering
    \resizebox{\columnwidth}{!}{%
        \input{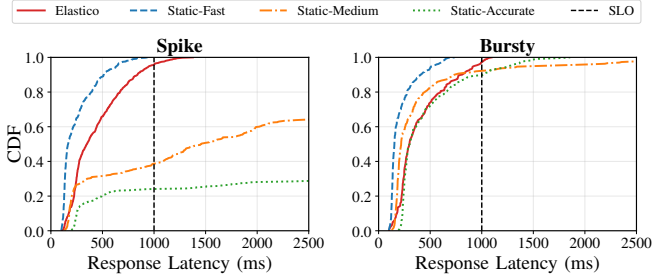}
    }
    \caption{Elastico latency CDF under 1000ms SLO.}
    \label{fig:cdf}
\end{figure}

The CDF plots (Figure~\ref{fig:cdf}) illustrate why static configurations fail under dynamic load. During the spike pattern, Static-Accurate latency distribution shows a long tail extending beyond 2500ms, with only 30\% of requests meeting the 1000ms SLO. Static-Medium similarly struggles with roughly 40\% compliance.

Elastico's latency distribution closely tracks Static-Fast in the low-latency region but achieves this while using more accurate configurations when load permits. The sharp rise at the SLO threshold indicates effective queue management, Elastico switches to faster configurations before queue buildup causes violations.

\subsubsection{Temporal Adaptation Behavior}

\begin{figure}[h]
    \centering
    \includegraphics[width=\columnwidth]{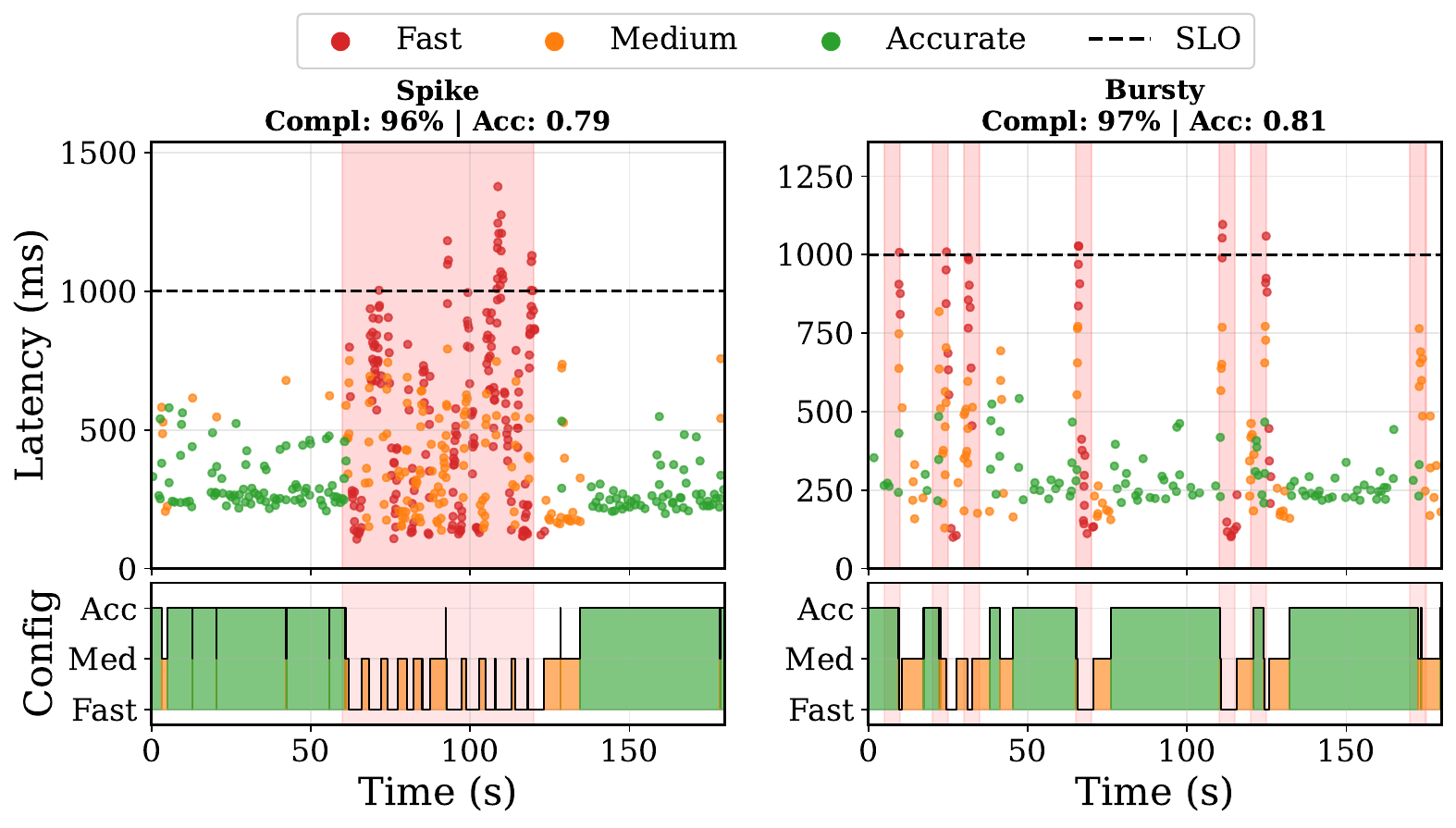}
    \caption{Elastico configuration switching over time under 1000ms SLO.}
    \label{fig:timeseries}
\end{figure}

The timeseries visualization (Figure~\ref{fig:timeseries}) demonstrates Elastico's adaptation mechanism in action. During the spike period (shaded region), Elastico rapidly transitions from the Accurate configuration to Fast, processing the load surge with minimal latency impact. As load subsides, Elastico gradually returns to more accurate configurations.

Key observations: (1) Configuration switches occur within seconds of load changes, demonstrating responsive adaptation; (2) During steady-state low load, Elastico favors accurate configurations (green points), maximizing answer quality; (3) During high load, Elastico predominantly uses fast configurations (red points), favoring SLO compliance.

%% file: figures/plots/experiments/compass_v/fig4_efficiency_scatter.pgf
%% Creator: Matplotlib, PGF backend
%%
%% To include the figure in your LaTeX document, write
%%   \input{<filename>.pgf}
%%
%% Make sure the required packages are loaded in your preamble
%%   \usepackage{pgf}
%%
%% Also ensure that all the required font packages are loaded; for instance,
%% the lmodern package is sometimes necessary when using math font.
%%   \usepackage{lmodern}
%%
%% Figures using additional raster images can only be included by \input if
%% they are in the same directory as the main LaTeX file. For loading figures
%% from other directories you can use the `import` package
%%   \usepackage{import}
%%
%% and then include the figures with
%%   \import{<path to file>}{<filename>.pgf}
%%
%% Matplotlib used the following preamble
%%   \def\mathdefault#1{#1}
%%   \everymath=\expandafter{\the\everymath\displaystyle}
%%   \IfFileExists{scrextend.sty}{
%%     \usepackage[fontsize=9.000000pt]{scrextend}
%%   }{
%%     \renewcommand{\normalsize}{\fontsize{9.000000}{10.800000}\selectfont}
%%     \normalsize
%%   }
%%   
%%   \makeatletter\@ifpackageloaded{underscore}{}{\usepackage[strings]{underscore}}\makeatother
%%
\begingroup%
\makeatletter%
\begin{pgfpicture}%
\pgfpathrectangle{\pgfpointorigin}{\pgfqpoint{4.625000in}{2.320369in}}%
\pgfusepath{use as bounding box, clip}%
\begin{pgfscope}%
\pgfsetbuttcap%
\pgfsetmiterjoin%
\definecolor{currentfill}{rgb}{1.000000,1.000000,1.000000}%
\pgfsetfillcolor{currentfill}%
\pgfsetlinewidth{0.000000pt}%
\definecolor{currentstroke}{rgb}{1.000000,1.000000,1.000000}%
\pgfsetstrokecolor{currentstroke}%
\pgfsetdash{}{0pt}%
\pgfpathmoveto{\pgfqpoint{0.000000in}{0.000000in}}%
\pgfpathlineto{\pgfqpoint{4.625000in}{0.000000in}}%
\pgfpathlineto{\pgfqpoint{4.625000in}{2.320369in}}%
\pgfpathlineto{\pgfqpoint{0.000000in}{2.320369in}}%
\pgfpathlineto{\pgfqpoint{0.000000in}{0.000000in}}%
\pgfpathclose%
\pgfusepath{fill}%
\end{pgfscope}%
\begin{pgfscope}%
\pgfsetbuttcap%
\pgfsetmiterjoin%
\definecolor{currentfill}{rgb}{1.000000,1.000000,1.000000}%
\pgfsetfillcolor{currentfill}%
\pgfsetlinewidth{0.000000pt}%
\definecolor{currentstroke}{rgb}{0.000000,0.000000,0.000000}%
\pgfsetstrokecolor{currentstroke}%
\pgfsetstrokeopacity{0.000000}%
\pgfsetdash{}{0pt}%
\pgfpathmoveto{\pgfqpoint{0.554864in}{0.476543in}}%
\pgfpathlineto{\pgfqpoint{4.525000in}{0.476543in}}%
\pgfpathlineto{\pgfqpoint{4.525000in}{1.998500in}}%
\pgfpathlineto{\pgfqpoint{0.554864in}{1.998500in}}%
\pgfpathlineto{\pgfqpoint{0.554864in}{0.476543in}}%
\pgfpathclose%
\pgfusepath{fill}%
\end{pgfscope}%
\begin{pgfscope}%
\pgfpathrectangle{\pgfqpoint{0.554864in}{0.476543in}}{\pgfqpoint{3.970136in}{1.521957in}}%
\pgfusepath{clip}%
\pgfsetrectcap%
\pgfsetroundjoin%
\pgfsetlinewidth{0.501875pt}%
\definecolor{currentstroke}{rgb}{0.690196,0.690196,0.690196}%
\pgfsetstrokecolor{currentstroke}%
\pgfsetstrokeopacity{0.200000}%
\pgfsetdash{}{0pt}%
\pgfpathmoveto{\pgfqpoint{0.629072in}{0.476543in}}%
\pgfpathlineto{\pgfqpoint{0.629072in}{1.998500in}}%
\pgfusepath{stroke}%
\end{pgfscope}%
\begin{pgfscope}%
\pgfsetbuttcap%
\pgfsetroundjoin%
\definecolor{currentfill}{rgb}{0.000000,0.000000,0.000000}%
\pgfsetfillcolor{currentfill}%
\pgfsetlinewidth{0.803000pt}%
\definecolor{currentstroke}{rgb}{0.000000,0.000000,0.000000}%
\pgfsetstrokecolor{currentstroke}%
\pgfsetdash{}{0pt}%
\pgfsys@defobject{currentmarker}{\pgfqpoint{0.000000in}{-0.048611in}}{\pgfqpoint{0.000000in}{0.000000in}}{%
\pgfpathmoveto{\pgfqpoint{0.000000in}{0.000000in}}%
\pgfpathlineto{\pgfqpoint{0.000000in}{-0.048611in}}%
\pgfusepath{stroke,fill}%
}%
\begin{pgfscope}%
\pgfsys@transformshift{0.629072in}{0.476543in}%
\pgfsys@useobject{currentmarker}{}%
\end{pgfscope}%
\end{pgfscope}%
\begin{pgfscope}%
\definecolor{textcolor}{rgb}{0.000000,0.000000,0.000000}%
\pgfsetstrokecolor{textcolor}%
\pgfsetfillcolor{textcolor}%
\pgftext[x=0.629072in,y=0.379321in,,top]{\color{textcolor}{\rmfamily\fontsize{8.000000}{9.600000}\selectfont\catcode`\^=\active\def^{\ifmmode\sp\else\^{}\fi}\catcode`\%=\active\def%{\%}$\mathdefault{0}$}}%
\end{pgfscope}%
\begin{pgfscope}%
\pgfpathrectangle{\pgfqpoint{0.554864in}{0.476543in}}{\pgfqpoint{3.970136in}{1.521957in}}%
\pgfusepath{clip}%
\pgfsetrectcap%
\pgfsetroundjoin%
\pgfsetlinewidth{0.501875pt}%
\definecolor{currentstroke}{rgb}{0.690196,0.690196,0.690196}%
\pgfsetstrokecolor{currentstroke}%
\pgfsetstrokeopacity{0.200000}%
\pgfsetdash{}{0pt}%
\pgfpathmoveto{\pgfqpoint{1.371153in}{0.476543in}}%
\pgfpathlineto{\pgfqpoint{1.371153in}{1.998500in}}%
\pgfusepath{stroke}%
\end{pgfscope}%
\begin{pgfscope}%
\pgfsetbuttcap%
\pgfsetroundjoin%
\definecolor{currentfill}{rgb}{0.000000,0.000000,0.000000}%
\pgfsetfillcolor{currentfill}%
\pgfsetlinewidth{0.803000pt}%
\definecolor{currentstroke}{rgb}{0.000000,0.000000,0.000000}%
\pgfsetstrokecolor{currentstroke}%
\pgfsetdash{}{0pt}%
\pgfsys@defobject{currentmarker}{\pgfqpoint{0.000000in}{-0.048611in}}{\pgfqpoint{0.000000in}{0.000000in}}{%
\pgfpathmoveto{\pgfqpoint{0.000000in}{0.000000in}}%
\pgfpathlineto{\pgfqpoint{0.000000in}{-0.048611in}}%
\pgfusepath{stroke,fill}%
}%
\begin{pgfscope}%
\pgfsys@transformshift{1.371153in}{0.476543in}%
\pgfsys@useobject{currentmarker}{}%
\end{pgfscope}%
\end{pgfscope}%
\begin{pgfscope}%
\definecolor{textcolor}{rgb}{0.000000,0.000000,0.000000}%
\pgfsetstrokecolor{textcolor}%
\pgfsetfillcolor{textcolor}%
\pgftext[x=1.371153in,y=0.379321in,,top]{\color{textcolor}{\rmfamily\fontsize{8.000000}{9.600000}\selectfont\catcode`\^=\active\def^{\ifmmode\sp\else\^{}\fi}\catcode`\%=\active\def%{\%}$\mathdefault{20}$}}%
\end{pgfscope}%
\begin{pgfscope}%
\pgfpathrectangle{\pgfqpoint{0.554864in}{0.476543in}}{\pgfqpoint{3.970136in}{1.521957in}}%
\pgfusepath{clip}%
\pgfsetrectcap%
\pgfsetroundjoin%
\pgfsetlinewidth{0.501875pt}%
\definecolor{currentstroke}{rgb}{0.690196,0.690196,0.690196}%
\pgfsetstrokecolor{currentstroke}%
\pgfsetstrokeopacity{0.200000}%
\pgfsetdash{}{0pt}%
\pgfpathmoveto{\pgfqpoint{2.113235in}{0.476543in}}%
\pgfpathlineto{\pgfqpoint{2.113235in}{1.998500in}}%
\pgfusepath{stroke}%
\end{pgfscope}%
\begin{pgfscope}%
\pgfsetbuttcap%
\pgfsetroundjoin%
\definecolor{currentfill}{rgb}{0.000000,0.000000,0.000000}%
\pgfsetfillcolor{currentfill}%
\pgfsetlinewidth{0.803000pt}%
\definecolor{currentstroke}{rgb}{0.000000,0.000000,0.000000}%
\pgfsetstrokecolor{currentstroke}%
\pgfsetdash{}{0pt}%
\pgfsys@defobject{currentmarker}{\pgfqpoint{0.000000in}{-0.048611in}}{\pgfqpoint{0.000000in}{0.000000in}}{%
\pgfpathmoveto{\pgfqpoint{0.000000in}{0.000000in}}%
\pgfpathlineto{\pgfqpoint{0.000000in}{-0.048611in}}%
\pgfusepath{stroke,fill}%
}%
\begin{pgfscope}%
\pgfsys@transformshift{2.113235in}{0.476543in}%
\pgfsys@useobject{currentmarker}{}%
\end{pgfscope}%
\end{pgfscope}%
\begin{pgfscope}%
\definecolor{textcolor}{rgb}{0.000000,0.000000,0.000000}%
\pgfsetstrokecolor{textcolor}%
\pgfsetfillcolor{textcolor}%
\pgftext[x=2.113235in,y=0.379321in,,top]{\color{textcolor}{\rmfamily\fontsize{8.000000}{9.600000}\selectfont\catcode`\^=\active\def^{\ifmmode\sp\else\^{}\fi}\catcode`\%=\active\def%{\%}$\mathdefault{40}$}}%
\end{pgfscope}%
\begin{pgfscope}%
\pgfpathrectangle{\pgfqpoint{0.554864in}{0.476543in}}{\pgfqpoint{3.970136in}{1.521957in}}%
\pgfusepath{clip}%
\pgfsetrectcap%
\pgfsetroundjoin%
\pgfsetlinewidth{0.501875pt}%
\definecolor{currentstroke}{rgb}{0.690196,0.690196,0.690196}%
\pgfsetstrokecolor{currentstroke}%
\pgfsetstrokeopacity{0.200000}%
\pgfsetdash{}{0pt}%
\pgfpathmoveto{\pgfqpoint{2.855316in}{0.476543in}}%
\pgfpathlineto{\pgfqpoint{2.855316in}{1.998500in}}%
\pgfusepath{stroke}%
\end{pgfscope}%
\begin{pgfscope}%
\pgfsetbuttcap%
\pgfsetroundjoin%
\definecolor{currentfill}{rgb}{0.000000,0.000000,0.000000}%
\pgfsetfillcolor{currentfill}%
\pgfsetlinewidth{0.803000pt}%
\definecolor{currentstroke}{rgb}{0.000000,0.000000,0.000000}%
\pgfsetstrokecolor{currentstroke}%
\pgfsetdash{}{0pt}%
\pgfsys@defobject{currentmarker}{\pgfqpoint{0.000000in}{-0.048611in}}{\pgfqpoint{0.000000in}{0.000000in}}{%
\pgfpathmoveto{\pgfqpoint{0.000000in}{0.000000in}}%
\pgfpathlineto{\pgfqpoint{0.000000in}{-0.048611in}}%
\pgfusepath{stroke,fill}%
}%
\begin{pgfscope}%
\pgfsys@transformshift{2.855316in}{0.476543in}%
\pgfsys@useobject{currentmarker}{}%
\end{pgfscope}%
\end{pgfscope}%
\begin{pgfscope}%
\definecolor{textcolor}{rgb}{0.000000,0.000000,0.000000}%
\pgfsetstrokecolor{textcolor}%
\pgfsetfillcolor{textcolor}%
\pgftext[x=2.855316in,y=0.379321in,,top]{\color{textcolor}{\rmfamily\fontsize{8.000000}{9.600000}\selectfont\catcode`\^=\active\def^{\ifmmode\sp\else\^{}\fi}\catcode`\%=\active\def%{\%}$\mathdefault{60}$}}%
\end{pgfscope}%
\begin{pgfscope}%
\pgfpathrectangle{\pgfqpoint{0.554864in}{0.476543in}}{\pgfqpoint{3.970136in}{1.521957in}}%
\pgfusepath{clip}%
\pgfsetrectcap%
\pgfsetroundjoin%
\pgfsetlinewidth{0.501875pt}%
\definecolor{currentstroke}{rgb}{0.690196,0.690196,0.690196}%
\pgfsetstrokecolor{currentstroke}%
\pgfsetstrokeopacity{0.200000}%
\pgfsetdash{}{0pt}%
\pgfpathmoveto{\pgfqpoint{3.597398in}{0.476543in}}%
\pgfpathlineto{\pgfqpoint{3.597398in}{1.998500in}}%
\pgfusepath{stroke}%
\end{pgfscope}%
\begin{pgfscope}%
\pgfsetbuttcap%
\pgfsetroundjoin%
\definecolor{currentfill}{rgb}{0.000000,0.000000,0.000000}%
\pgfsetfillcolor{currentfill}%
\pgfsetlinewidth{0.803000pt}%
\definecolor{currentstroke}{rgb}{0.000000,0.000000,0.000000}%
\pgfsetstrokecolor{currentstroke}%
\pgfsetdash{}{0pt}%
\pgfsys@defobject{currentmarker}{\pgfqpoint{0.000000in}{-0.048611in}}{\pgfqpoint{0.000000in}{0.000000in}}{%
\pgfpathmoveto{\pgfqpoint{0.000000in}{0.000000in}}%
\pgfpathlineto{\pgfqpoint{0.000000in}{-0.048611in}}%
\pgfusepath{stroke,fill}%
}%
\begin{pgfscope}%
\pgfsys@transformshift{3.597398in}{0.476543in}%
\pgfsys@useobject{currentmarker}{}%
\end{pgfscope}%
\end{pgfscope}%
\begin{pgfscope}%
\definecolor{textcolor}{rgb}{0.000000,0.000000,0.000000}%
\pgfsetstrokecolor{textcolor}%
\pgfsetfillcolor{textcolor}%
\pgftext[x=3.597398in,y=0.379321in,,top]{\color{textcolor}{\rmfamily\fontsize{8.000000}{9.600000}\selectfont\catcode`\^=\active\def^{\ifmmode\sp\else\^{}\fi}\catcode`\%=\active\def%{\%}$\mathdefault{80}$}}%
\end{pgfscope}%
\begin{pgfscope}%
\pgfpathrectangle{\pgfqpoint{0.554864in}{0.476543in}}{\pgfqpoint{3.970136in}{1.521957in}}%
\pgfusepath{clip}%
\pgfsetrectcap%
\pgfsetroundjoin%
\pgfsetlinewidth{0.501875pt}%
\definecolor{currentstroke}{rgb}{0.690196,0.690196,0.690196}%
\pgfsetstrokecolor{currentstroke}%
\pgfsetstrokeopacity{0.200000}%
\pgfsetdash{}{0pt}%
\pgfpathmoveto{\pgfqpoint{4.339480in}{0.476543in}}%
\pgfpathlineto{\pgfqpoint{4.339480in}{1.998500in}}%
\pgfusepath{stroke}%
\end{pgfscope}%
\begin{pgfscope}%
\pgfsetbuttcap%
\pgfsetroundjoin%
\definecolor{currentfill}{rgb}{0.000000,0.000000,0.000000}%
\pgfsetfillcolor{currentfill}%
\pgfsetlinewidth{0.803000pt}%
\definecolor{currentstroke}{rgb}{0.000000,0.000000,0.000000}%
\pgfsetstrokecolor{currentstroke}%
\pgfsetdash{}{0pt}%
\pgfsys@defobject{currentmarker}{\pgfqpoint{0.000000in}{-0.048611in}}{\pgfqpoint{0.000000in}{0.000000in}}{%
\pgfpathmoveto{\pgfqpoint{0.000000in}{0.000000in}}%
\pgfpathlineto{\pgfqpoint{0.000000in}{-0.048611in}}%
\pgfusepath{stroke,fill}%
}%
\begin{pgfscope}%
\pgfsys@transformshift{4.339480in}{0.476543in}%
\pgfsys@useobject{currentmarker}{}%
\end{pgfscope}%
\end{pgfscope}%
\begin{pgfscope}%
\definecolor{textcolor}{rgb}{0.000000,0.000000,0.000000}%
\pgfsetstrokecolor{textcolor}%
\pgfsetfillcolor{textcolor}%
\pgftext[x=4.339480in,y=0.379321in,,top]{\color{textcolor}{\rmfamily\fontsize{8.000000}{9.600000}\selectfont\catcode`\^=\active\def^{\ifmmode\sp\else\^{}\fi}\catcode`\%=\active\def%{\%}$\mathdefault{100}$}}%
\end{pgfscope}%
\begin{pgfscope}%
\definecolor{textcolor}{rgb}{0.000000,0.000000,0.000000}%
\pgfsetstrokecolor{textcolor}%
\pgfsetfillcolor{textcolor}%
\pgftext[x=2.539932in,y=0.225000in,,top]{\color{textcolor}{\rmfamily\fontsize{9.000000}{10.800000}\selectfont\catcode`\^=\active\def^{\ifmmode\sp\else\^{}\fi}\catcode`\%=\active\def%{\%}Feasible Configs (\%)}}%
\end{pgfscope}%
\begin{pgfscope}%
\pgfpathrectangle{\pgfqpoint{0.554864in}{0.476543in}}{\pgfqpoint{3.970136in}{1.521957in}}%
\pgfusepath{clip}%
\pgfsetrectcap%
\pgfsetroundjoin%
\pgfsetlinewidth{0.501875pt}%
\definecolor{currentstroke}{rgb}{0.690196,0.690196,0.690196}%
\pgfsetstrokecolor{currentstroke}%
\pgfsetstrokeopacity{0.200000}%
\pgfsetdash{}{0pt}%
\pgfpathmoveto{\pgfqpoint{0.554864in}{0.476543in}}%
\pgfpathlineto{\pgfqpoint{4.525000in}{0.476543in}}%
\pgfusepath{stroke}%
\end{pgfscope}%
\begin{pgfscope}%
\pgfsetbuttcap%
\pgfsetroundjoin%
\definecolor{currentfill}{rgb}{0.000000,0.000000,0.000000}%
\pgfsetfillcolor{currentfill}%
\pgfsetlinewidth{0.803000pt}%
\definecolor{currentstroke}{rgb}{0.000000,0.000000,0.000000}%
\pgfsetstrokecolor{currentstroke}%
\pgfsetdash{}{0pt}%
\pgfsys@defobject{currentmarker}{\pgfqpoint{-0.048611in}{0.000000in}}{\pgfqpoint{-0.000000in}{0.000000in}}{%
\pgfpathmoveto{\pgfqpoint{-0.000000in}{0.000000in}}%
\pgfpathlineto{\pgfqpoint{-0.048611in}{0.000000in}}%
\pgfusepath{stroke,fill}%
}%
\begin{pgfscope}%
\pgfsys@transformshift{0.554864in}{0.476543in}%
\pgfsys@useobject{currentmarker}{}%
\end{pgfscope}%
\end{pgfscope}%
\begin{pgfscope}%
\definecolor{textcolor}{rgb}{0.000000,0.000000,0.000000}%
\pgfsetstrokecolor{textcolor}%
\pgfsetfillcolor{textcolor}%
\pgftext[x=0.398613in, y=0.437963in, left, base]{\color{textcolor}{\rmfamily\fontsize{8.000000}{9.600000}\selectfont\catcode`\^=\active\def^{\ifmmode\sp\else\^{}\fi}\catcode`\%=\active\def%{\%}$\mathdefault{0}$}}%
\end{pgfscope}%
\begin{pgfscope}%
\pgfpathrectangle{\pgfqpoint{0.554864in}{0.476543in}}{\pgfqpoint{3.970136in}{1.521957in}}%
\pgfusepath{clip}%
\pgfsetrectcap%
\pgfsetroundjoin%
\pgfsetlinewidth{0.501875pt}%
\definecolor{currentstroke}{rgb}{0.690196,0.690196,0.690196}%
\pgfsetstrokecolor{currentstroke}%
\pgfsetstrokeopacity{0.200000}%
\pgfsetdash{}{0pt}%
\pgfpathmoveto{\pgfqpoint{0.554864in}{0.780935in}}%
\pgfpathlineto{\pgfqpoint{4.525000in}{0.780935in}}%
\pgfusepath{stroke}%
\end{pgfscope}%
\begin{pgfscope}%
\pgfsetbuttcap%
\pgfsetroundjoin%
\definecolor{currentfill}{rgb}{0.000000,0.000000,0.000000}%
\pgfsetfillcolor{currentfill}%
\pgfsetlinewidth{0.803000pt}%
\definecolor{currentstroke}{rgb}{0.000000,0.000000,0.000000}%
\pgfsetstrokecolor{currentstroke}%
\pgfsetdash{}{0pt}%
\pgfsys@defobject{currentmarker}{\pgfqpoint{-0.048611in}{0.000000in}}{\pgfqpoint{-0.000000in}{0.000000in}}{%
\pgfpathmoveto{\pgfqpoint{-0.000000in}{0.000000in}}%
\pgfpathlineto{\pgfqpoint{-0.048611in}{0.000000in}}%
\pgfusepath{stroke,fill}%
}%
\begin{pgfscope}%
\pgfsys@transformshift{0.554864in}{0.780935in}%
\pgfsys@useobject{currentmarker}{}%
\end{pgfscope}%
\end{pgfscope}%
\begin{pgfscope}%
\definecolor{textcolor}{rgb}{0.000000,0.000000,0.000000}%
\pgfsetstrokecolor{textcolor}%
\pgfsetfillcolor{textcolor}%
\pgftext[x=0.339584in, y=0.742354in, left, base]{\color{textcolor}{\rmfamily\fontsize{8.000000}{9.600000}\selectfont\catcode`\^=\active\def^{\ifmmode\sp\else\^{}\fi}\catcode`\%=\active\def%{\%}$\mathdefault{20}$}}%
\end{pgfscope}%
\begin{pgfscope}%
\pgfpathrectangle{\pgfqpoint{0.554864in}{0.476543in}}{\pgfqpoint{3.970136in}{1.521957in}}%
\pgfusepath{clip}%
\pgfsetrectcap%
\pgfsetroundjoin%
\pgfsetlinewidth{0.501875pt}%
\definecolor{currentstroke}{rgb}{0.690196,0.690196,0.690196}%
\pgfsetstrokecolor{currentstroke}%
\pgfsetstrokeopacity{0.200000}%
\pgfsetdash{}{0pt}%
\pgfpathmoveto{\pgfqpoint{0.554864in}{1.085326in}}%
\pgfpathlineto{\pgfqpoint{4.525000in}{1.085326in}}%
\pgfusepath{stroke}%
\end{pgfscope}%
\begin{pgfscope}%
\pgfsetbuttcap%
\pgfsetroundjoin%
\definecolor{currentfill}{rgb}{0.000000,0.000000,0.000000}%
\pgfsetfillcolor{currentfill}%
\pgfsetlinewidth{0.803000pt}%
\definecolor{currentstroke}{rgb}{0.000000,0.000000,0.000000}%
\pgfsetstrokecolor{currentstroke}%
\pgfsetdash{}{0pt}%
\pgfsys@defobject{currentmarker}{\pgfqpoint{-0.048611in}{0.000000in}}{\pgfqpoint{-0.000000in}{0.000000in}}{%
\pgfpathmoveto{\pgfqpoint{-0.000000in}{0.000000in}}%
\pgfpathlineto{\pgfqpoint{-0.048611in}{0.000000in}}%
\pgfusepath{stroke,fill}%
}%
\begin{pgfscope}%
\pgfsys@transformshift{0.554864in}{1.085326in}%
\pgfsys@useobject{currentmarker}{}%
\end{pgfscope}%
\end{pgfscope}%
\begin{pgfscope}%
\definecolor{textcolor}{rgb}{0.000000,0.000000,0.000000}%
\pgfsetstrokecolor{textcolor}%
\pgfsetfillcolor{textcolor}%
\pgftext[x=0.339584in, y=1.046746in, left, base]{\color{textcolor}{\rmfamily\fontsize{8.000000}{9.600000}\selectfont\catcode`\^=\active\def^{\ifmmode\sp\else\^{}\fi}\catcode`\%=\active\def%{\%}$\mathdefault{40}$}}%
\end{pgfscope}%
\begin{pgfscope}%
\pgfpathrectangle{\pgfqpoint{0.554864in}{0.476543in}}{\pgfqpoint{3.970136in}{1.521957in}}%
\pgfusepath{clip}%
\pgfsetrectcap%
\pgfsetroundjoin%
\pgfsetlinewidth{0.501875pt}%
\definecolor{currentstroke}{rgb}{0.690196,0.690196,0.690196}%
\pgfsetstrokecolor{currentstroke}%
\pgfsetstrokeopacity{0.200000}%
\pgfsetdash{}{0pt}%
\pgfpathmoveto{\pgfqpoint{0.554864in}{1.389717in}}%
\pgfpathlineto{\pgfqpoint{4.525000in}{1.389717in}}%
\pgfusepath{stroke}%
\end{pgfscope}%
\begin{pgfscope}%
\pgfsetbuttcap%
\pgfsetroundjoin%
\definecolor{currentfill}{rgb}{0.000000,0.000000,0.000000}%
\pgfsetfillcolor{currentfill}%
\pgfsetlinewidth{0.803000pt}%
\definecolor{currentstroke}{rgb}{0.000000,0.000000,0.000000}%
\pgfsetstrokecolor{currentstroke}%
\pgfsetdash{}{0pt}%
\pgfsys@defobject{currentmarker}{\pgfqpoint{-0.048611in}{0.000000in}}{\pgfqpoint{-0.000000in}{0.000000in}}{%
\pgfpathmoveto{\pgfqpoint{-0.000000in}{0.000000in}}%
\pgfpathlineto{\pgfqpoint{-0.048611in}{0.000000in}}%
\pgfusepath{stroke,fill}%
}%
\begin{pgfscope}%
\pgfsys@transformshift{0.554864in}{1.389717in}%
\pgfsys@useobject{currentmarker}{}%
\end{pgfscope}%
\end{pgfscope}%
\begin{pgfscope}%
\definecolor{textcolor}{rgb}{0.000000,0.000000,0.000000}%
\pgfsetstrokecolor{textcolor}%
\pgfsetfillcolor{textcolor}%
\pgftext[x=0.339584in, y=1.351137in, left, base]{\color{textcolor}{\rmfamily\fontsize{8.000000}{9.600000}\selectfont\catcode`\^=\active\def^{\ifmmode\sp\else\^{}\fi}\catcode`\%=\active\def%{\%}$\mathdefault{60}$}}%
\end{pgfscope}%
\begin{pgfscope}%
\pgfpathrectangle{\pgfqpoint{0.554864in}{0.476543in}}{\pgfqpoint{3.970136in}{1.521957in}}%
\pgfusepath{clip}%
\pgfsetrectcap%
\pgfsetroundjoin%
\pgfsetlinewidth{0.501875pt}%
\definecolor{currentstroke}{rgb}{0.690196,0.690196,0.690196}%
\pgfsetstrokecolor{currentstroke}%
\pgfsetstrokeopacity{0.200000}%
\pgfsetdash{}{0pt}%
\pgfpathmoveto{\pgfqpoint{0.554864in}{1.694109in}}%
\pgfpathlineto{\pgfqpoint{4.525000in}{1.694109in}}%
\pgfusepath{stroke}%
\end{pgfscope}%
\begin{pgfscope}%
\pgfsetbuttcap%
\pgfsetroundjoin%
\definecolor{currentfill}{rgb}{0.000000,0.000000,0.000000}%
\pgfsetfillcolor{currentfill}%
\pgfsetlinewidth{0.803000pt}%
\definecolor{currentstroke}{rgb}{0.000000,0.000000,0.000000}%
\pgfsetstrokecolor{currentstroke}%
\pgfsetdash{}{0pt}%
\pgfsys@defobject{currentmarker}{\pgfqpoint{-0.048611in}{0.000000in}}{\pgfqpoint{-0.000000in}{0.000000in}}{%
\pgfpathmoveto{\pgfqpoint{-0.000000in}{0.000000in}}%
\pgfpathlineto{\pgfqpoint{-0.048611in}{0.000000in}}%
\pgfusepath{stroke,fill}%
}%
\begin{pgfscope}%
\pgfsys@transformshift{0.554864in}{1.694109in}%
\pgfsys@useobject{currentmarker}{}%
\end{pgfscope}%
\end{pgfscope}%
\begin{pgfscope}%
\definecolor{textcolor}{rgb}{0.000000,0.000000,0.000000}%
\pgfsetstrokecolor{textcolor}%
\pgfsetfillcolor{textcolor}%
\pgftext[x=0.339584in, y=1.655528in, left, base]{\color{textcolor}{\rmfamily\fontsize{8.000000}{9.600000}\selectfont\catcode`\^=\active\def^{\ifmmode\sp\else\^{}\fi}\catcode`\%=\active\def%{\%}$\mathdefault{80}$}}%
\end{pgfscope}%
\begin{pgfscope}%
\pgfpathrectangle{\pgfqpoint{0.554864in}{0.476543in}}{\pgfqpoint{3.970136in}{1.521957in}}%
\pgfusepath{clip}%
\pgfsetrectcap%
\pgfsetroundjoin%
\pgfsetlinewidth{0.501875pt}%
\definecolor{currentstroke}{rgb}{0.690196,0.690196,0.690196}%
\pgfsetstrokecolor{currentstroke}%
\pgfsetstrokeopacity{0.200000}%
\pgfsetdash{}{0pt}%
\pgfpathmoveto{\pgfqpoint{0.554864in}{1.998500in}}%
\pgfpathlineto{\pgfqpoint{4.525000in}{1.998500in}}%
\pgfusepath{stroke}%
\end{pgfscope}%
\begin{pgfscope}%
\pgfsetbuttcap%
\pgfsetroundjoin%
\definecolor{currentfill}{rgb}{0.000000,0.000000,0.000000}%
\pgfsetfillcolor{currentfill}%
\pgfsetlinewidth{0.803000pt}%
\definecolor{currentstroke}{rgb}{0.000000,0.000000,0.000000}%
\pgfsetstrokecolor{currentstroke}%
\pgfsetdash{}{0pt}%
\pgfsys@defobject{currentmarker}{\pgfqpoint{-0.048611in}{0.000000in}}{\pgfqpoint{-0.000000in}{0.000000in}}{%
\pgfpathmoveto{\pgfqpoint{-0.000000in}{0.000000in}}%
\pgfpathlineto{\pgfqpoint{-0.048611in}{0.000000in}}%
\pgfusepath{stroke,fill}%
}%
\begin{pgfscope}%
\pgfsys@transformshift{0.554864in}{1.998500in}%
\pgfsys@useobject{currentmarker}{}%
\end{pgfscope}%
\end{pgfscope}%
\begin{pgfscope}%
\definecolor{textcolor}{rgb}{0.000000,0.000000,0.000000}%
\pgfsetstrokecolor{textcolor}%
\pgfsetfillcolor{textcolor}%
\pgftext[x=0.280556in, y=1.959920in, left, base]{\color{textcolor}{\rmfamily\fontsize{8.000000}{9.600000}\selectfont\catcode`\^=\active\def^{\ifmmode\sp\else\^{}\fi}\catcode`\%=\active\def%{\%}$\mathdefault{100}$}}%
\end{pgfscope}%
\begin{pgfscope}%
\definecolor{textcolor}{rgb}{0.000000,0.000000,0.000000}%
\pgfsetstrokecolor{textcolor}%
\pgfsetfillcolor{textcolor}%
\pgftext[x=0.225000in,y=1.237522in,,bottom,rotate=90.000000]{\color{textcolor}{\rmfamily\fontsize{9.000000}{10.800000}\selectfont\catcode`\^=\active\def^{\ifmmode\sp\else\^{}\fi}\catcode`\%=\active\def%{\%}Savings over Grid Search (\%)}}%
\end{pgfscope}%
\begin{pgfscope}%
\pgfsetrectcap%
\pgfsetmiterjoin%
\pgfsetlinewidth{0.803000pt}%
\definecolor{currentstroke}{rgb}{0.000000,0.000000,0.000000}%
\pgfsetstrokecolor{currentstroke}%
\pgfsetdash{}{0pt}%
\pgfpathmoveto{\pgfqpoint{0.554864in}{0.476543in}}%
\pgfpathlineto{\pgfqpoint{0.554864in}{1.998500in}}%
\pgfusepath{stroke}%
\end{pgfscope}%
\begin{pgfscope}%
\pgfsetrectcap%
\pgfsetmiterjoin%
\pgfsetlinewidth{0.803000pt}%
\definecolor{currentstroke}{rgb}{0.000000,0.000000,0.000000}%
\pgfsetstrokecolor{currentstroke}%
\pgfsetdash{}{0pt}%
\pgfpathmoveto{\pgfqpoint{4.525000in}{0.476543in}}%
\pgfpathlineto{\pgfqpoint{4.525000in}{1.998500in}}%
\pgfusepath{stroke}%
\end{pgfscope}%
\begin{pgfscope}%
\pgfsetrectcap%
\pgfsetmiterjoin%
\pgfsetlinewidth{0.803000pt}%
\definecolor{currentstroke}{rgb}{0.000000,0.000000,0.000000}%
\pgfsetstrokecolor{currentstroke}%
\pgfsetdash{}{0pt}%
\pgfpathmoveto{\pgfqpoint{0.554864in}{0.476543in}}%
\pgfpathlineto{\pgfqpoint{4.525000in}{0.476543in}}%
\pgfusepath{stroke}%
\end{pgfscope}%
\begin{pgfscope}%
\pgfsetrectcap%
\pgfsetmiterjoin%
\pgfsetlinewidth{0.803000pt}%
\definecolor{currentstroke}{rgb}{0.000000,0.000000,0.000000}%
\pgfsetstrokecolor{currentstroke}%
\pgfsetdash{}{0pt}%
\pgfpathmoveto{\pgfqpoint{0.554864in}{1.998500in}}%
\pgfpathlineto{\pgfqpoint{4.525000in}{1.998500in}}%
\pgfusepath{stroke}%
\end{pgfscope}%
\begin{pgfscope}%
\pgfpathrectangle{\pgfqpoint{0.554864in}{0.476543in}}{\pgfqpoint{3.970136in}{1.521957in}}%
\pgfusepath{clip}%
\pgfsetrectcap%
\pgfsetroundjoin%
\pgfsetlinewidth{1.204500pt}%
\definecolor{currentstroke}{rgb}{0.121569,0.466667,0.705882}%
\pgfsetstrokecolor{currentstroke}%
\pgfsetdash{}{0pt}%
\pgfpathmoveto{\pgfqpoint{0.706990in}{1.765641in}}%
\pgfpathlineto{\pgfqpoint{1.118846in}{1.482557in}}%
\pgfpathlineto{\pgfqpoint{1.849796in}{1.164468in}}%
\pgfpathlineto{\pgfqpoint{3.196674in}{0.785501in}}%
\pgfpathlineto{\pgfqpoint{3.656765in}{0.965091in}}%
\pgfpathlineto{\pgfqpoint{4.005543in}{1.207083in}}%
\pgfpathlineto{\pgfqpoint{4.306086in}{1.517562in}}%
\pgfusepath{stroke}%
\end{pgfscope}%
\begin{pgfscope}%
\pgfpathrectangle{\pgfqpoint{0.554864in}{0.476543in}}{\pgfqpoint{3.970136in}{1.521957in}}%
\pgfusepath{clip}%
\pgfsetbuttcap%
\pgfsetroundjoin%
\definecolor{currentfill}{rgb}{0.121569,0.466667,0.705882}%
\pgfsetfillcolor{currentfill}%
\pgfsetlinewidth{0.401500pt}%
\definecolor{currentstroke}{rgb}{1.000000,1.000000,1.000000}%
\pgfsetstrokecolor{currentstroke}%
\pgfsetdash{}{0pt}%
\pgfsys@defobject{currentmarker}{\pgfqpoint{-0.034722in}{-0.034722in}}{\pgfqpoint{0.034722in}{0.034722in}}{%
\pgfpathmoveto{\pgfqpoint{0.000000in}{-0.034722in}}%
\pgfpathcurveto{\pgfqpoint{0.009208in}{-0.034722in}}{\pgfqpoint{0.018041in}{-0.031064in}}{\pgfqpoint{0.024552in}{-0.024552in}}%
\pgfpathcurveto{\pgfqpoint{0.031064in}{-0.018041in}}{\pgfqpoint{0.034722in}{-0.009208in}}{\pgfqpoint{0.034722in}{0.000000in}}%
\pgfpathcurveto{\pgfqpoint{0.034722in}{0.009208in}}{\pgfqpoint{0.031064in}{0.018041in}}{\pgfqpoint{0.024552in}{0.024552in}}%
\pgfpathcurveto{\pgfqpoint{0.018041in}{0.031064in}}{\pgfqpoint{0.009208in}{0.034722in}}{\pgfqpoint{0.000000in}{0.034722in}}%
\pgfpathcurveto{\pgfqpoint{-0.009208in}{0.034722in}}{\pgfqpoint{-0.018041in}{0.031064in}}{\pgfqpoint{-0.024552in}{0.024552in}}%
\pgfpathcurveto{\pgfqpoint{-0.031064in}{0.018041in}}{\pgfqpoint{-0.034722in}{0.009208in}}{\pgfqpoint{-0.034722in}{0.000000in}}%
\pgfpathcurveto{\pgfqpoint{-0.034722in}{-0.009208in}}{\pgfqpoint{-0.031064in}{-0.018041in}}{\pgfqpoint{-0.024552in}{-0.024552in}}%
\pgfpathcurveto{\pgfqpoint{-0.018041in}{-0.031064in}}{\pgfqpoint{-0.009208in}{-0.034722in}}{\pgfqpoint{0.000000in}{-0.034722in}}%
\pgfpathlineto{\pgfqpoint{0.000000in}{-0.034722in}}%
\pgfpathclose%
\pgfusepath{stroke,fill}%
}%
\begin{pgfscope}%
\pgfsys@transformshift{0.706990in}{1.765641in}%
\pgfsys@useobject{currentmarker}{}%
\end{pgfscope}%
\begin{pgfscope}%
\pgfsys@transformshift{1.118846in}{1.482557in}%
\pgfsys@useobject{currentmarker}{}%
\end{pgfscope}%
\begin{pgfscope}%
\pgfsys@transformshift{1.849796in}{1.164468in}%
\pgfsys@useobject{currentmarker}{}%
\end{pgfscope}%
\begin{pgfscope}%
\pgfsys@transformshift{3.196674in}{0.785501in}%
\pgfsys@useobject{currentmarker}{}%
\end{pgfscope}%
\begin{pgfscope}%
\pgfsys@transformshift{3.656765in}{0.965091in}%
\pgfsys@useobject{currentmarker}{}%
\end{pgfscope}%
\begin{pgfscope}%
\pgfsys@transformshift{4.005543in}{1.207083in}%
\pgfsys@useobject{currentmarker}{}%
\end{pgfscope}%
\begin{pgfscope}%
\pgfsys@transformshift{4.306086in}{1.517562in}%
\pgfsys@useobject{currentmarker}{}%
\end{pgfscope}%
\end{pgfscope}%
\begin{pgfscope}%
\pgfpathrectangle{\pgfqpoint{0.554864in}{0.476543in}}{\pgfqpoint{3.970136in}{1.521957in}}%
\pgfusepath{clip}%
\pgfsetrectcap%
\pgfsetroundjoin%
\pgfsetlinewidth{1.204500pt}%
\definecolor{currentstroke}{rgb}{0.878431,0.482353,0.223529}%
\pgfsetstrokecolor{currentstroke}%
\pgfsetdash{}{0pt}%
\pgfpathmoveto{\pgfqpoint{0.996402in}{1.598225in}}%
\pgfpathlineto{\pgfqpoint{1.285814in}{1.496254in}}%
\pgfpathlineto{\pgfqpoint{2.094683in}{1.385151in}}%
\pgfpathlineto{\pgfqpoint{2.721742in}{1.289268in}}%
\pgfpathlineto{\pgfqpoint{3.482375in}{1.254263in}}%
\pgfpathlineto{\pgfqpoint{3.990701in}{1.319707in}}%
\pgfpathlineto{\pgfqpoint{4.194774in}{1.511474in}}%
\pgfpathlineto{\pgfqpoint{4.291244in}{1.683455in}}%
\pgfusepath{stroke}%
\end{pgfscope}%
\begin{pgfscope}%
\pgfpathrectangle{\pgfqpoint{0.554864in}{0.476543in}}{\pgfqpoint{3.970136in}{1.521957in}}%
\pgfusepath{clip}%
\pgfsetbuttcap%
\pgfsetmiterjoin%
\definecolor{currentfill}{rgb}{0.878431,0.482353,0.223529}%
\pgfsetfillcolor{currentfill}%
\pgfsetlinewidth{0.401500pt}%
\definecolor{currentstroke}{rgb}{1.000000,1.000000,1.000000}%
\pgfsetstrokecolor{currentstroke}%
\pgfsetdash{}{0pt}%
\pgfsys@defobject{currentmarker}{\pgfqpoint{-0.049105in}{-0.049105in}}{\pgfqpoint{0.049105in}{0.049105in}}{%
\pgfpathmoveto{\pgfqpoint{-0.000000in}{-0.049105in}}%
\pgfpathlineto{\pgfqpoint{0.049105in}{0.000000in}}%
\pgfpathlineto{\pgfqpoint{0.000000in}{0.049105in}}%
\pgfpathlineto{\pgfqpoint{-0.049105in}{0.000000in}}%
\pgfpathlineto{\pgfqpoint{-0.000000in}{-0.049105in}}%
\pgfpathclose%
\pgfusepath{stroke,fill}%
}%
\begin{pgfscope}%
\pgfsys@transformshift{0.996402in}{1.598225in}%
\pgfsys@useobject{currentmarker}{}%
\end{pgfscope}%
\begin{pgfscope}%
\pgfsys@transformshift{1.285814in}{1.496254in}%
\pgfsys@useobject{currentmarker}{}%
\end{pgfscope}%
\begin{pgfscope}%
\pgfsys@transformshift{2.094683in}{1.385151in}%
\pgfsys@useobject{currentmarker}{}%
\end{pgfscope}%
\begin{pgfscope}%
\pgfsys@transformshift{2.721742in}{1.289268in}%
\pgfsys@useobject{currentmarker}{}%
\end{pgfscope}%
\begin{pgfscope}%
\pgfsys@transformshift{3.482375in}{1.254263in}%
\pgfsys@useobject{currentmarker}{}%
\end{pgfscope}%
\begin{pgfscope}%
\pgfsys@transformshift{3.990701in}{1.319707in}%
\pgfsys@useobject{currentmarker}{}%
\end{pgfscope}%
\begin{pgfscope}%
\pgfsys@transformshift{4.194774in}{1.511474in}%
\pgfsys@useobject{currentmarker}{}%
\end{pgfscope}%
\begin{pgfscope}%
\pgfsys@transformshift{4.291244in}{1.683455in}%
\pgfsys@useobject{currentmarker}{}%
\end{pgfscope}%
\end{pgfscope}%
\begin{pgfscope}%
\pgfsetbuttcap%
\pgfsetmiterjoin%
\definecolor{currentfill}{rgb}{1.000000,1.000000,1.000000}%
\pgfsetfillcolor{currentfill}%
\pgfsetfillopacity{0.900000}%
\pgfsetlinewidth{1.003750pt}%
\definecolor{currentstroke}{rgb}{0.800000,0.800000,0.800000}%
\pgfsetstrokecolor{currentstroke}%
\pgfsetstrokeopacity{0.900000}%
\pgfsetdash{}{0pt}%
\pgfpathmoveto{\pgfqpoint{1.087264in}{2.039813in}}%
\pgfpathlineto{\pgfqpoint{3.992599in}{2.039813in}}%
\pgfpathquadraticcurveto{\pgfqpoint{4.013433in}{2.039813in}}{\pgfqpoint{4.013433in}{2.060647in}}%
\pgfpathlineto{\pgfqpoint{4.013433in}{2.199536in}}%
\pgfpathquadraticcurveto{\pgfqpoint{4.013433in}{2.220369in}}{\pgfqpoint{3.992599in}{2.220369in}}%
\pgfpathlineto{\pgfqpoint{1.087264in}{2.220369in}}%
\pgfpathquadraticcurveto{\pgfqpoint{1.066431in}{2.220369in}}{\pgfqpoint{1.066431in}{2.199536in}}%
\pgfpathlineto{\pgfqpoint{1.066431in}{2.060647in}}%
\pgfpathquadraticcurveto{\pgfqpoint{1.066431in}{2.039813in}}{\pgfqpoint{1.087264in}{2.039813in}}%
\pgfpathlineto{\pgfqpoint{1.087264in}{2.039813in}}%
\pgfpathclose%
\pgfusepath{stroke,fill}%
\end{pgfscope}%
\begin{pgfscope}%
\pgfsetrectcap%
\pgfsetroundjoin%
\pgfsetlinewidth{1.204500pt}%
\definecolor{currentstroke}{rgb}{0.121569,0.466667,0.705882}%
\pgfsetstrokecolor{currentstroke}%
\pgfsetdash{}{0pt}%
\pgfpathmoveto{\pgfqpoint{1.108098in}{2.142244in}}%
\pgfpathlineto{\pgfqpoint{1.212264in}{2.142244in}}%
\pgfpathlineto{\pgfqpoint{1.316431in}{2.142244in}}%
\pgfusepath{stroke}%
\end{pgfscope}%
\begin{pgfscope}%
\pgfsetbuttcap%
\pgfsetroundjoin%
\definecolor{currentfill}{rgb}{0.121569,0.466667,0.705882}%
\pgfsetfillcolor{currentfill}%
\pgfsetlinewidth{0.401500pt}%
\definecolor{currentstroke}{rgb}{1.000000,1.000000,1.000000}%
\pgfsetstrokecolor{currentstroke}%
\pgfsetdash{}{0pt}%
\pgfsys@defobject{currentmarker}{\pgfqpoint{-0.034722in}{-0.034722in}}{\pgfqpoint{0.034722in}{0.034722in}}{%
\pgfpathmoveto{\pgfqpoint{0.000000in}{-0.034722in}}%
\pgfpathcurveto{\pgfqpoint{0.009208in}{-0.034722in}}{\pgfqpoint{0.018041in}{-0.031064in}}{\pgfqpoint{0.024552in}{-0.024552in}}%
\pgfpathcurveto{\pgfqpoint{0.031064in}{-0.018041in}}{\pgfqpoint{0.034722in}{-0.009208in}}{\pgfqpoint{0.034722in}{0.000000in}}%
\pgfpathcurveto{\pgfqpoint{0.034722in}{0.009208in}}{\pgfqpoint{0.031064in}{0.018041in}}{\pgfqpoint{0.024552in}{0.024552in}}%
\pgfpathcurveto{\pgfqpoint{0.018041in}{0.031064in}}{\pgfqpoint{0.009208in}{0.034722in}}{\pgfqpoint{0.000000in}{0.034722in}}%
\pgfpathcurveto{\pgfqpoint{-0.009208in}{0.034722in}}{\pgfqpoint{-0.018041in}{0.031064in}}{\pgfqpoint{-0.024552in}{0.024552in}}%
\pgfpathcurveto{\pgfqpoint{-0.031064in}{0.018041in}}{\pgfqpoint{-0.034722in}{0.009208in}}{\pgfqpoint{-0.034722in}{0.000000in}}%
\pgfpathcurveto{\pgfqpoint{-0.034722in}{-0.009208in}}{\pgfqpoint{-0.031064in}{-0.018041in}}{\pgfqpoint{-0.024552in}{-0.024552in}}%
\pgfpathcurveto{\pgfqpoint{-0.018041in}{-0.031064in}}{\pgfqpoint{-0.009208in}{-0.034722in}}{\pgfqpoint{0.000000in}{-0.034722in}}%
\pgfpathlineto{\pgfqpoint{0.000000in}{-0.034722in}}%
\pgfpathclose%
\pgfusepath{stroke,fill}%
}%
\begin{pgfscope}%
\pgfsys@transformshift{1.212264in}{2.142244in}%
\pgfsys@useobject{currentmarker}{}%
\end{pgfscope}%
\end{pgfscope}%
\begin{pgfscope}%
\definecolor{textcolor}{rgb}{0.000000,0.000000,0.000000}%
\pgfsetstrokecolor{textcolor}%
\pgfsetfillcolor{textcolor}%
\pgftext[x=1.347681in,y=2.105786in,left,base]{\color{textcolor}{\rmfamily\fontsize{7.500000}{9.000000}\selectfont\catcode`\^=\active\def^{\ifmmode\sp\else\^{}\fi}\catcode`\%=\active\def%{\%}RAG (234 configs)}}%
\end{pgfscope}%
\begin{pgfscope}%
\pgfsetrectcap%
\pgfsetroundjoin%
\pgfsetlinewidth{1.204500pt}%
\definecolor{currentstroke}{rgb}{0.878431,0.482353,0.223529}%
\pgfsetstrokecolor{currentstroke}%
\pgfsetdash{}{0pt}%
\pgfpathmoveto{\pgfqpoint{2.353492in}{2.142244in}}%
\pgfpathlineto{\pgfqpoint{2.457658in}{2.142244in}}%
\pgfpathlineto{\pgfqpoint{2.561825in}{2.142244in}}%
\pgfusepath{stroke}%
\end{pgfscope}%
\begin{pgfscope}%
\pgfsetbuttcap%
\pgfsetmiterjoin%
\definecolor{currentfill}{rgb}{0.878431,0.482353,0.223529}%
\pgfsetfillcolor{currentfill}%
\pgfsetlinewidth{0.401500pt}%
\definecolor{currentstroke}{rgb}{1.000000,1.000000,1.000000}%
\pgfsetstrokecolor{currentstroke}%
\pgfsetdash{}{0pt}%
\pgfsys@defobject{currentmarker}{\pgfqpoint{-0.049105in}{-0.049105in}}{\pgfqpoint{0.049105in}{0.049105in}}{%
\pgfpathmoveto{\pgfqpoint{-0.000000in}{-0.049105in}}%
\pgfpathlineto{\pgfqpoint{0.049105in}{0.000000in}}%
\pgfpathlineto{\pgfqpoint{0.000000in}{0.049105in}}%
\pgfpathlineto{\pgfqpoint{-0.049105in}{0.000000in}}%
\pgfpathlineto{\pgfqpoint{-0.000000in}{-0.049105in}}%
\pgfpathclose%
\pgfusepath{stroke,fill}%
}%
\begin{pgfscope}%
\pgfsys@transformshift{2.457658in}{2.142244in}%
\pgfsys@useobject{currentmarker}{}%
\end{pgfscope}%
\end{pgfscope}%
\begin{pgfscope}%
\definecolor{textcolor}{rgb}{0.000000,0.000000,0.000000}%
\pgfsetstrokecolor{textcolor}%
\pgfsetfillcolor{textcolor}%
\pgftext[x=2.593075in,y=2.105786in,left,base]{\color{textcolor}{\rmfamily\fontsize{7.500000}{9.000000}\selectfont\catcode`\^=\active\def^{\ifmmode\sp\else\^{}\fi}\catcode`\%=\active\def%{\%}Obj.\ Detection (385 configs)}}%
\end{pgfscope}%
\end{pgfpicture}%
\makeatother%
\endgroup%

%% file: sections/related_work.tex
\section{Related Work}
\label{sec:related_work}

% We review prior work across three areas relevant to dynamic adaptation in Compound AI systems. First, we analyze recent optimization approaches for Compound AI workflows. Next, we examine model selection and serving systems that handle model selection as an adaptation mechanism for inference workloads. Finally, While existing work addresses aspects of these challenges independently, none provides integrated solutions for dynamic adaptation in multi-model workflows.

We review prior work across three areas relevant to Compound AI dynamic adaptation: (1) optimization of compound workflows, (2) model selection for inference serving, and (3) resource management for multi-stage pipelines.

\subsection{Compound AI Optimization}

Recent work addresses the challenge of optimizing Compound AI workflow quality~\cite{OPTSURVEY}. LLMSelector~\cite{chen2025llmselector} demonstrates that assigning different models to different workflow components outperforms uniform model allocation, using an LLM evaluator to learn per-component model strengths through iterative optimization. Optimas~\cite{optimas} extends this direction by maintaining local reward functions for each component that align with global system performance, enabling joint optimization of prompts, hyperparameters, and model parameters. Murakkab~\cite{MURAKKAB} addresses serving of agentic workflows in cloud platforms, using a profile-guided optimizer and declarative abstraction to map workflow components to models and hardware.

These approaches treat the optimized workflow as a static artifact. Once a configuration is selected and deployed, it remains fixed during serving. Compass builds on these optimization insights and extends them, enabling the system to (1) discover and (2) transition between pre-computed configurations as load conditions change.

\subsection{Model Selection and Serving}

Dynamic model selection is an established technique for balancing accuracy and latency in inference serving~\cite{CLIPPER, INFaaS, RAMSIS}. Clipper~\cite{CLIPPER} introduced the inference serving architecture, employing bandit algorithms to select efficient models per query. Model-Switching~\cite{switching} introduced scaling models vertically rather than scaling resources horizontally during load fluctuations, switching between model variants to maintain SLO compliance. RAMSIS~\cite{RAMSIS} advances model selection by formulating it as a Markov decision process (MDP), exploiting query inter-arrival patterns to maximize accuracy under latency constraints. Jellyfish~\cite{Jellyfish} applies similar principles to edge networks, adapting DNN selection based on network bandwidth dynamics.

These systems target scenarios where a single model serves each request. Compound AI workflows involve multiple models, where the output of one component affects the input size and latency of downstream components. Applying these techniques to individual components in isolation does not account for such interactions. Compass extends model selection principles to multi-component workflows, enabling dynamic adaptation through configuration switching. 

% These systems assume a single model per request, whereas Compound AI workflows involve multiple models and components whose outputs affect overall performance. Optimizing components in isolation ignores these interactions. 

\subsection{Compound AI Serving and Resource Management}

Closest to our work are systems that explicitly address serving multi-stage AI workflows. InferLine~\cite{InferLine} minimizes end-to-end latency by managing tail latencies and provisioning resources for individual pipeline stages. Circinus~\cite{liuCircinusEfficientQuery2025} optimizes operator placement and configuration across edge-cloud infrastructure tiers, using Bayesian optimization to efficiently search the configuration space for multi-query deployments.

InferLine relies on horizontal scaling and assumes elastic infrastructure where resources can be provisioned on demand. Circinus addresses a different problem: finding optimal static placements across heterogeneous tiers, triggering re-planning when conditions change. Neither system exploits the accuracy-latency trade-off through model variant selection as an adaptation mechanism. Compass targets constrained environments where scaling out is not possible. Instead of adapting infrastructure to fit the model, Compass adapts the Compound AI configuration to fit the infrastructure.

% In contrast, Compass targets constrained environments (e.g., dedicated inference infrastructure) where scaling out is not possible. Instead of adapting the infrastructure to fit the model (provisioning), Compass adapts the Compound AI configuration to fit the infrastructure (configuration switching). 

%% file: sections/conclusion.tex
\section{Conclusion}
\label{sec:conclusion}

This paper demonstrates that accuracy-latency trade-offs inherent in Compound AI workflows can be utilized to provide a practical mechanism for runtime adaptation through configuration switching. Compass discovers accuracy-feasible configurations that span a Pareto front of latency-accuracy operating points, enabling the selection of faster configurations under high load and more accurate configurations under low load, maintaining SLO compliance without horizontal scaling.

Compass separates the problem into two phases. The offline phase uses the COMPASS-V algorithm to discover feasible configurations and the Planner to derive switching policies through an analytical queuing-theory based model (AQM). The online phase uses the Elastico Controller to select configurations based on queue depth thresholds derived from these policies. Across two Compound AI workflows (RAG and multi-model object detection), COMPASS-V achieves 100\% recall in finding feasible configurations, while reducing evaluations by 57.5\% on average compared to exhaustive search, with efficiency gains reaching 95.3\% at tight accuracy thresholds. At runtime, Elastico improves SLO compliance by 71.6\% over static high-accuracy baselines, while simultaneously improving accuracy by 3-5 percentage points over static fast baselines, achieving 90-98\% SLO compliance across various workload patterns. 

Several directions remain open. First, the current implementation assumes all Pareto-optimal configurations fit in GPU memory on a single server. Extending Compass to multi-server deployments would require jointly deciding when to switch configurations versus when to add replicas, with cost and energy as first-class objectives. Next, the AQM assumes Poisson arrivals and reacts to load changes after they occur. Replacing the reactive model with predictive adaptation could enable anticipatory switching before queue buildup causes SLO violations. Finally, Compass switches the entire workflow configuration atomically. Modeling per-component contributions to overall workflow performance could enable finer-grained switching for more precise adaptation.